\documentclass[%
aip,
amsmath,amssymb,
reprint,%
]{revtex4-1}
\usepackage{graphicx}
\usepackage{dcolumn}
\usepackage{bm}
\usepackage{amsmath}
\usepackage{xspace}
\usepackage{xcolor}
\usepackage[utf8]{inputenc}
\usepackage[T1]{fontenc}
\usepackage{mathptmx}
\usepackage{listings}
\usepackage{adjustbox}
\usepackage[normalem]{ulem}
\definecolor{codegreen}{rgb}{0,0.6,0}
\definecolor{codeblue}{rgb}{0.,0.,1.}
\definecolor{codegray}{rgb}{0.5,0.5,0.5}
\definecolor{codepurple}{rgb}{0.58,0,0.82}
\definecolor{backcolour}{RGB}{248,248,250}
\lstdefinestyle{mystyle}{
	language=Python,
	backgroundcolor=\color{backcolour},   
	commentstyle=\color{codegreen},
	keywordstyle=\color{magenta},
	numberstyle=\tiny\color{codegray},
	stringstyle=\color{codepurple},
	basicstyle=\linespread{1.}\ttfamily\small,
	breakatwhitespace=true,         
	breaklines=true,                 
	captionpos=b,                    
	keepspaces=true,                 
	numbers=none,                    
	numbersep=5pt,                  
	showspaces=false,                
	showstringspaces=false,
	showtabs=false,                
	tabsize=4
}
\usepackage{comment}
\usepackage{multirow,booktabs,caption,threeparttable}
\newcommand{\pyscf}{\textsc{PySCF}\xspace}

\newcommand{\numpy}{\textsc{Numpy}\xspace}
\newcommand{\scipy}{\textsc{Scipy}\xspace}

\usepackage{color}
\newcommand{\green}[1]{{\color[rgb]{0,0.6,0}{#1}}}
\newcommand{\red}[1]{{\color[rgb]{0,0,0}{#1}}}
\newcommand{\blue}[1]{{\color[rgb]{0,0,0}{#1}}}
\newcommand{\tcb}[1]{\green{[TCB: #1]}}

\begin{document}
\title{
    Recent developments in the PySCF program package
}

\author{Qiming Sun            }   \affiliation{AxiomQuant Investment Management LLC, Shanghai, 200120, China} 
\author{Xing Zhang            }   \affiliation{Division of Chemistry and Chemical Engineering, California Institute of Technology, Pasadena, CA 91125, USA} 
\author{Samragni Banerjee     }   \affiliation{Department of Chemistry and Biochemistry, The Ohio State University, Columbus, OH 43210, USA} 
\author{Peng Bao              }   \affiliation{Beijing National Laboratory for Molecular Sciences, State Key Laboratory for Structural Chemistry of Unstable and Stable Species, Institute of Chemistry, Chinese Academy of Sciences, Beijing 100190, China} 
\author{Marc Barbry           }   \affiliation{Simbeyond B.V., P.O. Box 513, NL-5600 MB Eindhoven, The Netherlands} 
\author{Nick S. Blunt         }   \affiliation{Department of Chemistry, Lensfield Road, Cambridge, CB2 1EW, United Kingdom}
\author{Nikolay A. Bogdanov   }   \affiliation{Max Planck Institute for Solid State Research, Heisenbergstraße 1, 70569 Stuttgart, Germany} 
\author{George H. Booth       }   \affiliation{Department of Physics, King’s College London, Strand, London WC2R 2LS, United Kingdom}
\author{Jia Chen              }   \affiliation{Department of Physics, University of Florida, Gainesville, FL 32611, USA} 
                                  \affiliation{Quantum Theory Project, University of Florida, Gainesville, FL 32611, USA}
\author{Zhi-Hao Cui           }   \affiliation{Division of Chemistry and Chemical Engineering, California Institute of Technology, Pasadena, CA 91125, USA} 
\author{Janus Juul Eriksen    }   \affiliation{School of Chemistry, University of Bristol, Cantock’s Close, Bristol BS8 1TS, United Kingdom} 
\author{Yang Gao              }   \affiliation{Division of Engineering and Applied Science, California Institute of Technology, Pasadena, CA 91125, USA} 
\author{Sheng Guo             }   \affiliation{Google Inc., Mountain View, CA 94043, USA} 
\author{Jan Hermann           }   \affiliation{FU Berlin, Department of Mathematics and Computer Science, Arnimallee 6, 14195 Berlin, Germany} 
                                  \affiliation{TU Berlin, Machine Learning Group, Marchstr. 23, 10587 Berlin, Germany}
\author{Matthew R. Hermes     }   \affiliation{Department of Chemistry, Chemical Theory Center, and Supercomputing Institute, University of Minnesota, 207 Pleasant Street SE, Minneapolis, MN 55455, USA} 
\author{Kevin Koh             }   \affiliation{Department of Chemistry and Biochemistry, The University of Notre Dame du Lac, 251 Nieuwland Science Hall, Notre Dame, IN 46556, USA} 
\author{Peter Koval           }   \affiliation{Simune Atomistics S.L., Avenida Tolosa 76, Donostia-San Sebastian, Spain} 
\author{Susi Lehtola          }   \affiliation{Department of Chemistry, University of Helsinki, P.O. Box 55 (A. I. Virtasen aukio 1), FI-00014 Helsinki, Finland.} 
\author{Zhendong Li           }   \affiliation{Key Laboratory of Theoretical and Computational Photochemistry, Ministry of Education, College of Chemistry, Beijing Normal University, Beijing 100875, China}
\author{Junzi Liu             }   \affiliation{Department of Chemistry, The Johns Hopkins University, Baltimore, MD 21218, USA}
\author{Narbe Mardirossian    }   \affiliation{AMGEN Research, One Amgen Center Drive, Thousand Oaks, CA 91320, USA} 
\author{James D. McClain      }   \affiliation{DRW Holdings LLC, Chicago, IL 60661, USA} 
\author{Mario Motta           }   \affiliation{IBM Almaden Research Center, San Jose, CA 95120, USA} 
\author{Bastien Mussard       }   \affiliation{Department of Chemistry, University of Colorado, Boulder, CO 80302, USA} 
\author{Hung Q. Pham          }   \affiliation{Department of Chemistry, Chemical Theory Center, and Supercomputing Institute, University of Minnesota, 207 Pleasant Street SE, Minneapolis, MN 55455, USA} 
\author{Artem Pulkin          }   \affiliation{QuTech and Kavli Institute of Nanoscience, Delft University of Technology, The Netherlands} 
\author{Wirawan Purwanto      }   \affiliation{Information Technology Services, Old Dominion University, Norfolk, VA 23529, USA}
\author{Paul J. Robinson      }   \affiliation{Department of Chemistry, Columbia  University, New York, NY 10027, USA} 
\author{Enrico Ronca          }   \affiliation{Istituto per i Processi Chimico Fisici del CNR (IPCF-CNR), Via G. Moruzzi, 1, 56124, Pisa, Italy}
\author{Elvira Sayfutyarova   }   \affiliation{Department of Chemistry, Yale University, 225 Prospect Street, New Haven, CT 06520, USA} 
\author{Maximilian Scheurer   }   \affiliation{Interdisciplinary Center for Scientific Computing, Ruprecht-Karls University of Heidelberg, 205 Im Neuenheimer Feld, 69120 Heidelberg, Germany} 
\author{Henry F. Schurkus     }   \affiliation{Division of Chemistry and Chemical Engineering, California Institute of Technology, Pasadena, CA 91125, USA} 
\author{James E. T. Smith     }   \affiliation{Department of Chemistry, University of Colorado, Boulder, CO 80302, USA} 
\author{Chong Sun             }   \affiliation{Division of Chemistry and Chemical Engineering, California Institute of Technology, Pasadena, CA 91125, USA} 
\author{Shi-Ning Sun          }   \affiliation{Division of Engineering and Applied Science, California Institute of Technology, Pasadena, CA 91125, USA} 
\author{Shiv Upadhyay         }   \affiliation{Department of Chemistry, University of Pittsburgh, Pittsburgh, PA 15260} 
\author{Lucas K. Wagner       }   \affiliation{Department of Physics and Institute for Condensed Matter Theory, University of Illinois at Urbana-Champaign, IL 61801, USA} 
\author{Xiao Wang             }   \affiliation{Center for Computational Quantum Physics, Flatiron Institute, New York, NY 10010, USA} 
\author{Alec White            }   \affiliation{Division of Chemistry and Chemical Engineering, California Institute of Technology, Pasadena, CA 91125, USA} 
\author{James Daniel Whitfield}   \affiliation{Department of Physics and Astronomy, Dartmouth College, Hanover, NH 03755, USA} 
\author{Mark J. Williamson    }   \affiliation{Department of Chemistry, University of Cambridge, Lensfield Road, Cambridge CB2 1EW, United Kingdom} 
\author{Sebastian Wouters     }   \affiliation{Bricsys NV, Bellevue 5/201, 9050 Gent, Belgium}
\author{Jun Yang              }   \affiliation{Department of Chemistry, The University of Hong Kong, Pokfulam Road, Hong Kong SAR, China}
\author{Jason M. Yu           }   \affiliation{Department of Chemistry, University of California, Irvine, 1102 Natural Sciences II, Irvine, CA 92697-2025, USA} 
\author{Tianyu Zhu            }   \affiliation{Division of Chemistry and Chemical Engineering, California Institute of Technology, Pasadena, CA 91125, USA} 
\author{Timothy C. Berkelbach }   \affiliation{Department of Chemistry, Columbia University, New York, NY 10027, USA} 
                                  \affiliation{Center for Computational Quantum Physics, Flatiron Institute, New York, NY 10010, USA}
\author{Sandeep Sharma        }   \affiliation{Department of Chemistry, University of Colorado, Boulder, CO 80302, USA} 
\author{Alexander Yu. Sokolov }   \affiliation{Department of Chemistry and Biochemistry, The Ohio State University, Columbus, OH 43210, USA} 
\author{Garnet Kin-Lic Chan  } \email{gkc1000@gmail.com}  \affiliation{Division of Chemistry and Chemical Engineering, California Institute of Technology, Pasadena, CA 91125, USA} 

\begin{abstract}
	\noindent
	\pyscf~is a Python-based general-purpose electronic structure platform
	that both supports first-principles simulations of molecules and solids,
        as well as accelerates the development of new methodology and
	complex computational workflows. The present paper explains the design and philosophy behind \pyscf
        that enables it to meet these twin objectives.
	With several case studies, we show how users can easily implement their own methods using \pyscf as a development environment.
	We then summarize the capabilities of \pyscf for molecular and solid-state simulations.
	Finally, we describe the growing ecosystem of projects that use \pyscf across the domains of 
	quantum chemistry, materials science, machine learning and quantum information science.
	
\end{abstract}
\maketitle

\section{Introduction}

\label{sec:intro}
This article describes the current status of the Python Simulations of Chemistry Framework, also
known as \pyscf, as of version 1.7.1. The \pyscf project was originally started in 2014 by Sun, then in the group of Chan, in the
context of developing a tool to enable ab initio quantum embedding calculations. 
However, it rapidly outgrew its rather specialized roots to become a general purpose development platform
for quantum simulations and electronic structure theory. 
The early history of \pyscf is recounted in Ref.~\onlinecite{Sun2018}. 
Now, \pyscf is a production ready
tool that implements many of the most commonly used methods in molecular quantum chemistry and solid-state electronic structure. 
Since its inception, \pyscf has been a free and open-source package hosted on Github,\red{\cite{pyscfgit}} and is now also available
through pip,\red{\cite{pyscfpypi}} conda,\red{\cite{pyscfconda}} and a number of other distribution platforms.
It has a userbase numbering in the hundreds, and over 60 code contributors.
Beyond  chemistry and materials science, it has also found use in the areas of data science,\red{\cite{Chen2019,Lu2019}}
machine learning,\red{\cite{Dick2019,Ji2018,Hermann2019,Han2019,scfinitguess,Pfau2019,Choo2019}}
and quantum computing,\red{\cite{McClean2017,qiskit,Yamazaki2018}}
in both academia as well as in industry.
To mark its transition from a code developed by a single group to a broader community effort, the leadership of \pyscf
was expanded in 2019 to a board of directors.\red{\cite{pyscfweb}} 

While the fields of quantum chemistry and solid-state electronic structure are rich with excellent software,\red{\cite{Valiev2010,Furche2014,Shao2015,Parrish2017,Kresse1996,Joubert1999,Enkovaara2010,Giannozzi2017}}
the development of PySCF is guided by some unique principles. In order of priority:
\begin{enumerate}
\item \pyscf should be more than a computational tool; it should be a development platform.
  We aim for users to be empowered to modify the code, implement their own methods without the assistance
  of the original developers, and incorporate parts of the code in a modular fashion into their own projects;
\item Unlike many packages which focus on either molecular chemistry or materials science applications,
  \pyscf should support both equally, to allow
  calculations on molecules and materials to be carried out in the same numerical framework and with the same
  theoretical approximations;
\item \pyscf should enable users outside of the chemical sciences (such as workers in machine learning
  and quantum information theory) to carry out quantum chemistry simulations. 
\end{enumerate}

In the rest of this article, we elaborate on these guiding principles of \pyscf, describing how they have impacted the program design
and implementation and how they can be used to implement new functionality in new projects.  We provide a brief summary of the implemented methods
and conclude with an overview of the \pyscf ecosystem in different areas of science.

\section{The design philosophy behind \pyscf}

All quantum simulation workflows naturally require some level of programming and
customization. 
This may arise in simple tasks, such as scanning a potential energy surface, tabulating
results, or automating input generation, or in more advanced use cases that
include more substantial programming, such as
with complex data processing, incorporating logic into the computational workflow,
or when embedding customized algorithms into the computation. In either case,
the ability to program with and extend one's simulation software greatly empowers the user.
\pyscf is designed to serve as a basic program library that can facilitate custom computational
tasks and workflows, as well as form the starting point for the development of new
algorithms. 

To enable this, \pyscf is constructed as a library of modular components with a  loosely coupled structure.
The modules provide easily reusable functions,
with (where possible) simple implementations, 
and hooks are provided within the code to enable extensibility.
Optimized and competitive performance is, as much as possible, separated
out into a small number of lower level components which do not need to be
touched by the user. We elaborate on these design choices below:
\begin{itemize}
  \item \emph{Reusable functions for individual suboperations}.

    It is becoming common practice to provide a Python scripting interface for input and simulation control. 
    However, \pyscf goes beyond this by providing
    a rich set of Python APIs not only for the simulation models,
    but also for many of the individual sub-operations that compose the algorithms. For example,
    after input parsing, a mean-field Hartree-Fock (HF) or density functional theory (DFT) algorithm comprises many steps, including integral
    generation, guess initialization, assembling components of the Fock matrix and
    diagonalizing, and accelerating iterations to self-consistent convergence. All of these suboperations are exposed
    as \pyscf APIs, enabling one to rebuild or modify the self-consistent algorithm at will. Similarly, APIs are
    exposed for other essential components of electronic structure algorithms, such as integral transformations,
    density fitting, Hamiltonian manipulation, various many-electron and Green's functions solvers, computation of derivatives, relativistic
    corrections, and so forth, in essence across all the functionality of \pyscf.
    The package provides a large number of examples to demonstrate how these APIs can be used in customized calculations
    or methodology development.


    With at most some simple initialization statements, the \pyscf APIs can be executed at any place and in any order
    within a code without side-effects. This means that when implementing or extending the code, the user
    does not need to retain information on the program state, and can focus on the physical theory of interest. 
    For instance, using the above example, one can call the function to build a Fock matrix from
a given density matrix anywhere in the code, regardless of whether the density matrix in question is related to
a larger simulation. From a programming design perspective, this is because within \pyscf
no implicit global variables are used and functions are implemented free of side effects (or
with minimal side effects) in a largely functional programming style.
The \pyscf function APIs generally follow the \numpy/\scipy API style. In this
convention, the input arguments are simple Python built-in datatypes or \numpy arrays, avoiding
the need to understand complex objects and structures.

  \item \emph{Simple implementations}.

Python is amongst the simplest of the widely-used programming languages and is
the main implementation language in \pyscf. 
Apart from a few performance critical functions, over 90\% of \pyscf is written in Python, with dependencies on only
a small number of common external Python libraries (\numpy, \scipy, \textsc{h5py}).

Implementation language does not hide organizational complexity, however,
and structural simplicity in \pyscf is achieved via additional design
choices. In particular, \pyscf uses a mixed object oriented/functional paradigm: 
complex simulation data (e.g. data on the molecular geometry or cell parameters) and simulation models (e.g. whether a
mean-field calculation is a HF or DFT one) are organized in an object oriented style,
while individual function implementations follow a functional programming paradigm.
Deep object inheritance is rarely used. Unlike packages where external input configuration files
are used to control a simulation, the simulation parameters are simply held in the member variables of the simulation model object.

Where possible, \pyscf provides multiple implementations
of the same algorithm with the same API: one is designed to be easy to read and simple
to modify, and another is for optimized performance.
For example, the full configuration interaction module contains both 
a slower but simpler implementation as well as heavily optimized implementations, specialized
for specific Hamiltonian symmetries and spin types. The
optimized algorithms  have components that are written in C. This dual level
of implementation mimics the Python convention of having modules in both pure Python and C
with the same API (such as the \textsc{profile} and \textsc{cProfile} modules of the Python standard library).
It also reflects
the \pyscf development cycle, where often a simple reference Python implementation
is first produced before being further optimized.


  \item \emph{Easily modified runtime functionality}.

    In customized simulations, it is often necessary to modify the underlying
    functionality of a package. This can be complicated in a compiled program due to
    the need to consider detailed types and compilation dependencies across modules.
    In contrast, many parts of \pyscf are easy to modify both
    due to the design of \pyscf as well as the dynamic runtime resolution of methods and
    ``duck typing''  of Python.
Generally speaking, one can modify
functionality in one part of the code without needing to worry about
breaking other parts of the package. For example, one can modify the HF 
module with a custom Hamiltonian without considering whether it will work
in a DFT calculation; the program will continue to run so long as the computational task
involves HF and post-HF methods. Further, Python ``monkey patching'' (replacing functionality at runtime)
means that core \pyscf routines can be overwritten without even touching the code base of the library.

  \item \emph{Competitive performance}.
  
In many simulations, performance is still the critical consideration. 
This is typically the reason for implementing code in compiled languages such as \textsc{Fortran} or C/C++.
In \pyscf, the performance gap between Python and compiled languages is partly removed by 
a heavy reliance on \numpy and \scipy, which provide Python APIs to optimized algorithms
written in compiled languages.
Additional optimization is achieved in \pyscf
with custom C implementations where necessary. 
Performance critical spots, which occur primarily in the integral and tensor
operations, are implemented in C and heavily optimized. The use of additional C libraries
also allows us to achieve thread-level parallelism \red{via OpenMP}, bypassing Python's intrinsic multithreading limitations.
Since a simulation can often spend over 99\% of its runtime in the C libraries, the overhead
due to the remaining Python code is negligible. The combination of Python with C libraries
ensures \pyscf achieves  leading performance in many simulations.
\end{itemize}

\section{A common framework for molecules and crystalline materials}

\label{sec:crystalline}
Electronic structure packages typically focus on either molecular or materials simulations, and are
thus built around numerical approximations adapted to either case.
A central goal of \pyscf is to enable molecules and materials to be simulated with common numerical
approximations and theoretical models. Originally, \pyscf started as a Gaussian atomic orbital (AO) molecular code, and was
subsequently extended to enable simulations in a crystalline Gaussian basis.
Much of the seemingly new functionality required in a crystalline materials simulation is in fact analogous
to functionality in a molecular implementation, such as
\begin{enumerate}
\item Using a Bloch basis. In \pyscf we use a crystalline Gaussian AO basis,
  which is analogous to a symmetry adapted molecular AO basis;
\item Exploiting translational symmetry by enforcing momentum conservation. This is analogous to handling
  molecular point group symmetries;
\item Handling complex numbers, given that matrix elements
  between Bloch functions are generally complex. This is analogous to the requirements of a molecular calculation
  with complex orbitals. 
  \end{enumerate}
Other modifications are unique to the crystalline material setting, including:
\begin{enumerate}
\item  Techniques to handle divergences associated with the long-ranged nature of the Coulomb interaction, since the classical electron-electron,
   electron-nuclear, and nuclear-nuclear interactions are separately divergent.
   In \pyscf this is handled via the density fitting integral routines (see below) and by evaluating
   certain contributions using Ewald summation techniques;
\item Numerical techniques special to periodic functions, such as the fast Fourier transform (FFT), as well
  as approximations tailored to plane-wave implementations, such as certain pseudopotentials. \pyscf supports
  mixed crystalline  Gaussian and plane-wave expressions, using both analytic integrals
  as well as FFT on grids;
\item Techniques to accelerate convergence to the thermodynamic limit. In \pyscf, such corrections are
   implemented at the mean-field level by modifying the treatment of the exchange energy, 
   which is the leading finite-size correction.
\item Additional crystal lattice symmetries. 
  Currently \pyscf contains only experimental support for additional lattice symmetries.
  \end{enumerate}


In \pyscf, we identify the three-index density fitted integrals
as the central computational intermediate that allows us to largely unify
molecular and crystalline implementations.
This is because:
\begin{enumerate}
\item three-center ``density-fitted'' Gaussian integrals are key  to  fast implementations;
\item The use of the FFT to evaluate the potential of a pair density of AO functions, which is needed in 
  fast DFT implementations with pseudopotentials,\cite{VandeVondele2005} is formally equivalent to density fitting with
  plane-waves; 
\item The density fitted integrals can be adjusted to remove the Coulomb divergences in materials;\cite{McClain2017}
\item three-index Coulomb intermediates are sufficiently compact that they can be computed even in the crystalline setting.
\end{enumerate}

\pyscf provides a unified density fitting API for both molecules and
crystalline materials. In molecules, the auxiliary basis is assumed to be Gaussian AOs,
while in the periodic setting, different types of auxiliary bases are provided, including
plane-wave functions (in the FFTDF module), crystalline Gaussian AOs (in the GDF module) and
mixed plane-wave-Gaussian functions (in the MDF module).\cite{Sun2017}  
Different auxiliary bases are provided in periodic calculations as they are suited to different AO basis sets:
FFTDF is efficient for smooth AO functions when used with pseudopotentials; GDF is
more efficient for compact AO functions; and MDF 
allows a high accuracy treatment of the Coulomb problem regardless of the compactness of the underlying
atomic orbital basis.



\begin{figure*}
	\includegraphics{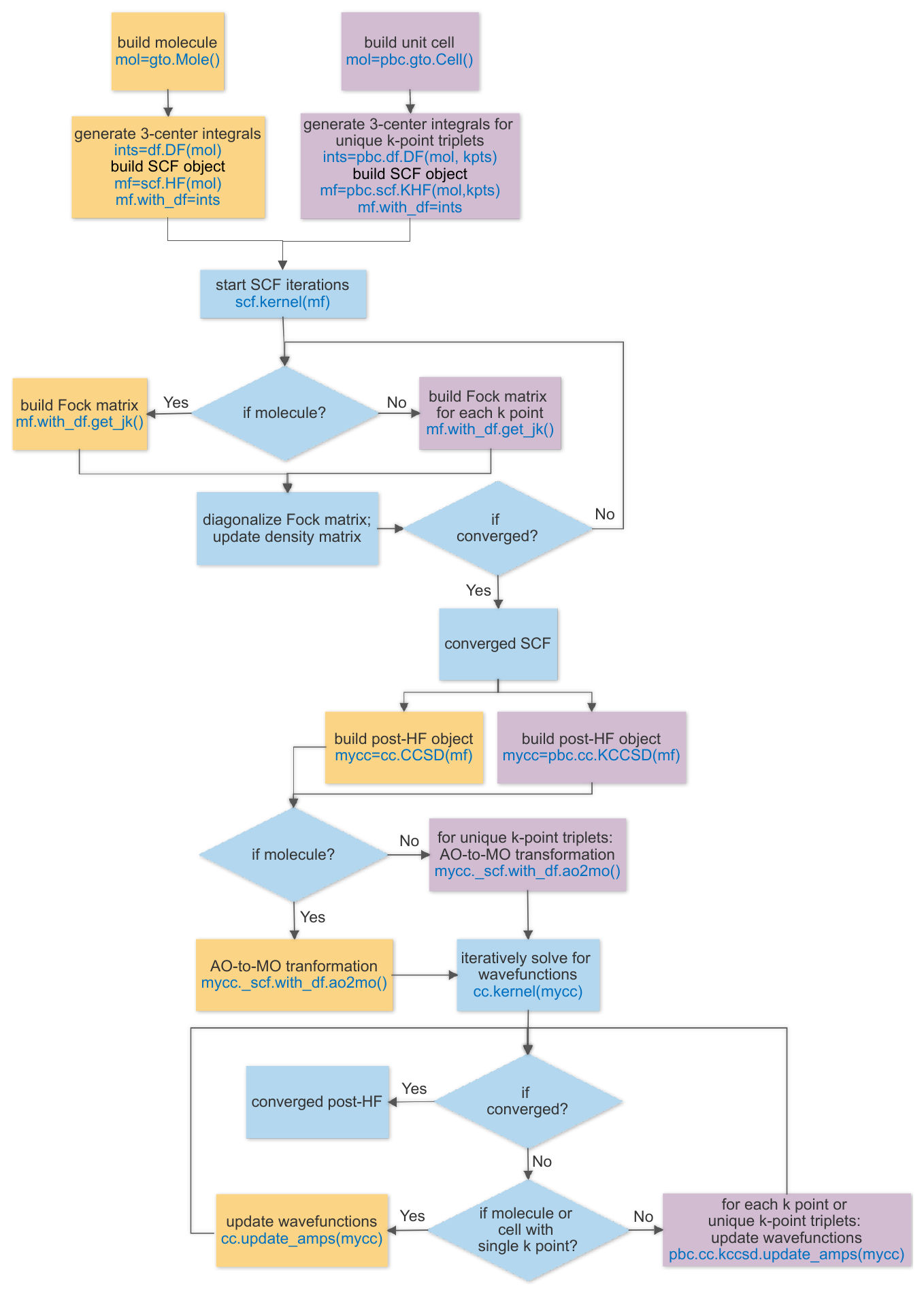}
	\caption{Illustration of the program workflow for molecular and periodic calculations.
		 The orange and purple boxes indicate functions that are $k$-point independent 
		 and $k$-point dependent, respectively; the blue boxes indicate generic driver
		 functions that can be used in both molecular and periodic calculations.
	}
	\label{fig:flowchart}
\end{figure*}

Using the above ideas, the general program structure, implementation, and simulation
workflow for molecular and materials calculations become very similar.
Figure \ref{fig:flowchart} shows an example of the computational workflow adopted in \pyscf for performing 
molecular and periodic post-HF calculations. 
The same driver functions can be used to carry out
generic operations such as solving the HF equations or coupled cluster amplitude equations.
However, the implementations of methods for molecular and crystalline systems 
bifurcate when evaluating $k$-point dependent quantities, 
such as the three-center density-fitted integrals, Hamiltonians, and wavefunctions. 
Nevertheless, if only a single $k$-point is considered (and especially at the $\Gamma$ point where
all integrals are real), most molecular modules can be used to 
perform calculations in crystals without modification (see Sec.~\ref{sec:methods}).

\section{Developing with \pyscf: case studies}

In this section we walk through some case studies that illustrate how the
functionality of \pyscf can be modified and extended. We focus on cases
which might be encountered by the average user who does not want to 
modify the source code, but wishes to assemble different existing \pyscf APIs to implement new functionality. 

\subsection{Case study: modifying the Hamiltonian}

\begin{figure*}
\includegraphics{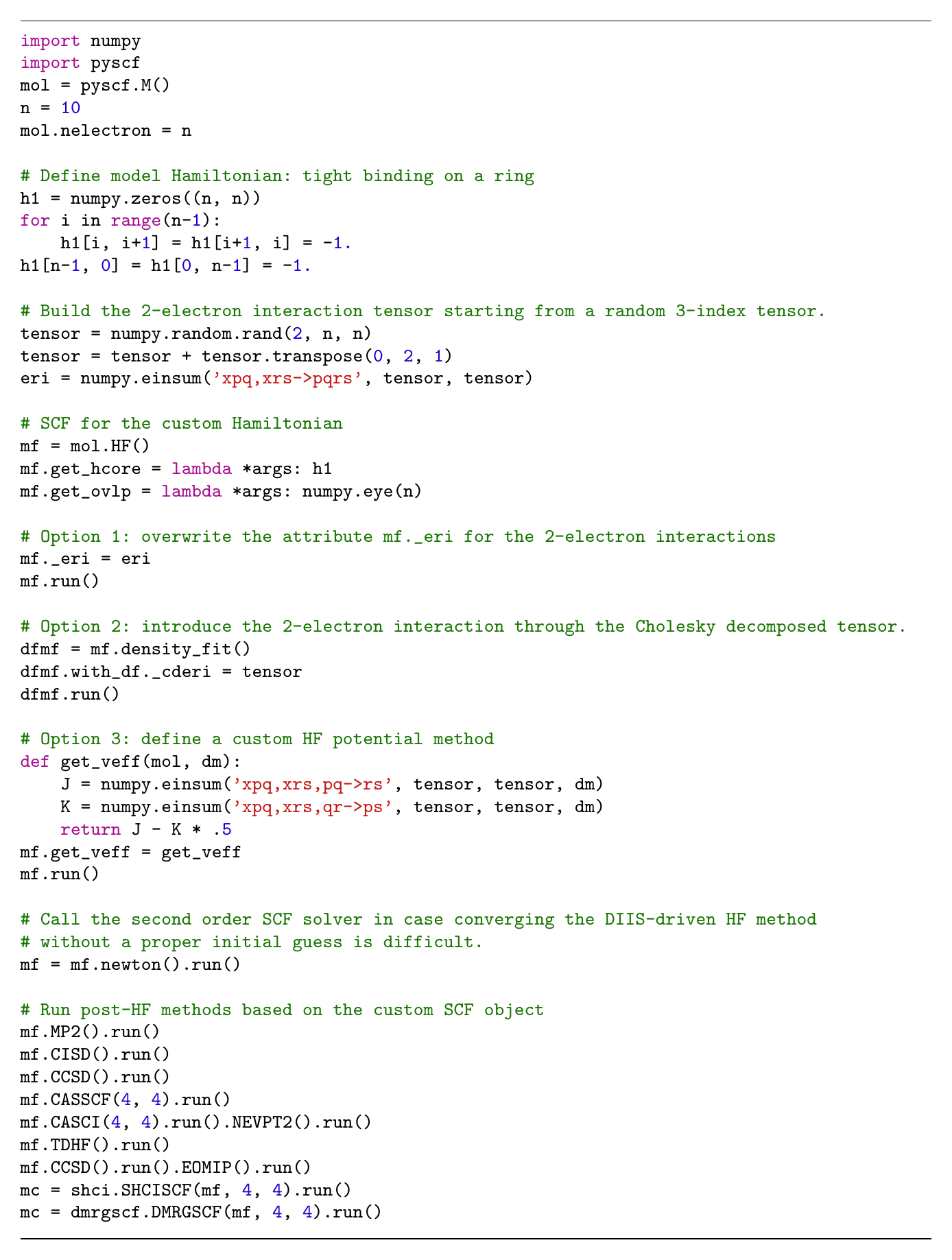}
  \caption{Hamiltonian customization and post-HF methods for customized Hamiltonians.
    }
  \label{code:customh2e}
\end{figure*}

In \pyscf, simulation models (i.e.~different wavefunction approximations)
are always implemented such that they can be used independently of any specific Hamiltonian,
with up to two-body interactions.
Consequently, the Hamiltonian under study can be easily customized by the user, 
which is useful for studying model problems or, for example, when trying to interface to
different numerical basis approximations.
Figure \ref{code:customh2e} shows several different ways to define
one-electron and two-electron interactions in the Hamiltonian 
followed by subsequent ground and excited state calculations with the custom Hamiltonian.
Note that if a method is not compatible with or well defined using the customized interactions, for instance,
in the case of solvation corrections, 
\pyscf will raise a Python runtime error in the place where the requisite operations are ill-defined.



\subsection{Case study: optimizing orbitals of arbitrary methods}

\label{sec:orbital_opt}
The \pyscf MCSCF module provides a general purpose quasi-second order orbital optimization
algorithm within orbital subspaces (e.g. active spaces) as well as over the complete orbital space.
In particular, it is not limited to the built-in CASCI, CASSCF and multi-reference correlation solvers, but allows orbital
  optimization of any method that provides energies   and one- and two-particle density matrices.
  For this reason, \pyscf is often used to carry out active space orbital optimization for DMRG (density matrix renormalization group), selected
  configuration interaction, and full configuration interaction quantum Monte Carlo wavefunctions, via its
  native interfaces to Block\cite{Sharma2012} (DMRG), CheMPS2\cite{Wouters2014} (DMRG), Dice\cite{Sharma2017,Smith2017,Holmes2016} (selected CI),
  Arrow\cite{Holmes2016,Sharma2017,Li2018} (selected CI), 
  and NECI\cite{Booth2014} (FCIQMC). 

\begin{figure*}
  \centering
  \includegraphics{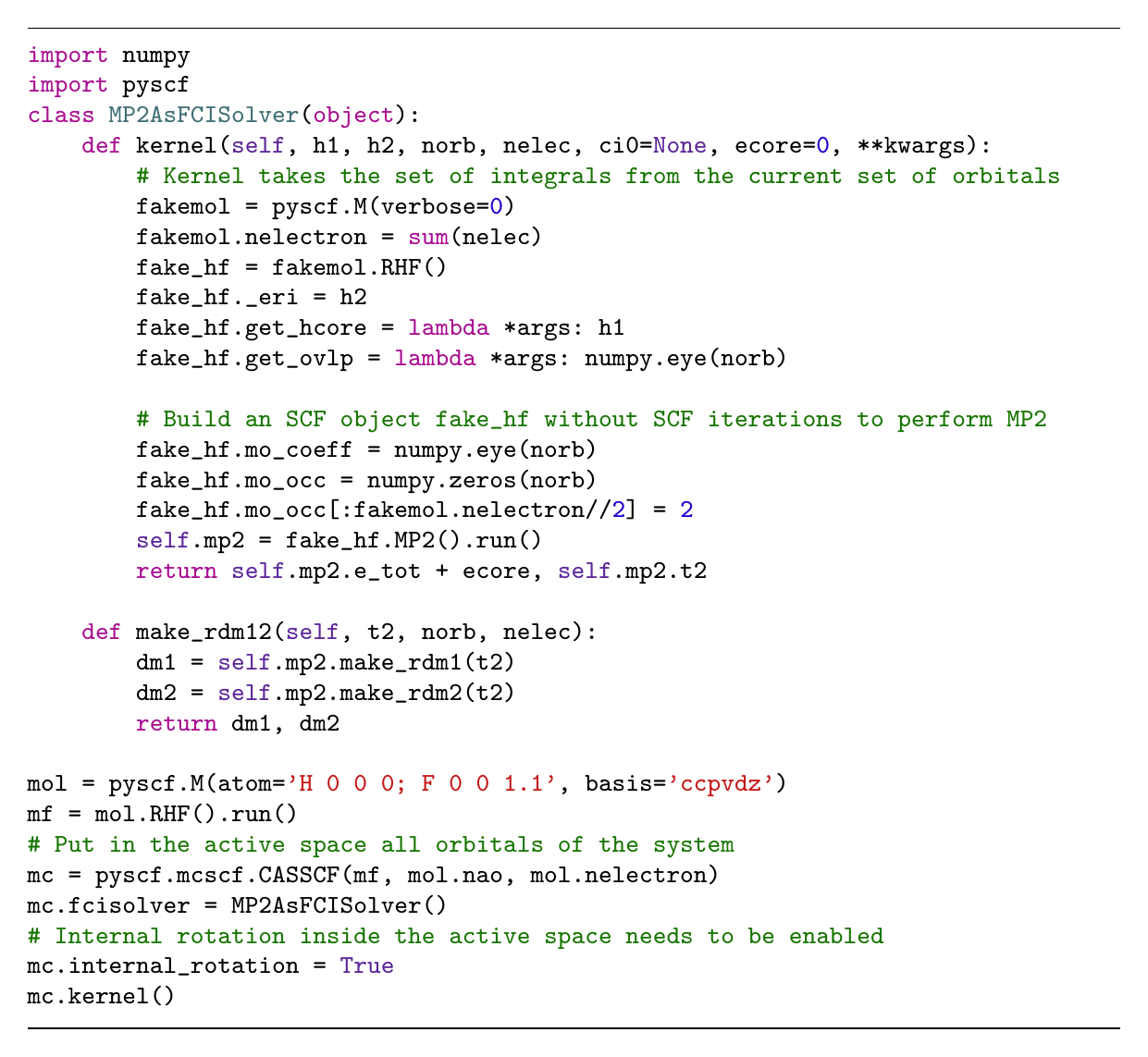}
  \caption{Using the general CASSCF solver to implement an orbital-optimized MP2 method.}
  \label{code:omp2}
\end{figure*}

  In addition, it is easy for the user to use the MCSCF module to optimize orbitals 
  in electronic structure methods for which the orbital optimization API is not natively implemented.
  For example, although orbital-optimized MP2\cite{Bozkaya2011} is not explicitly
  provided in \pyscf, a simple version of it can easily be performed using a short script, shown in
  Figure \ref{code:omp2}. Without any modifications, the orbital optimization will use
  a quasi-second order algorithm. We see that the user only needs to write a simple wrapper
  to provide  two functions, namely, \textsf{make\_rdm12}, 
  which computes the one- and two-particle density matrices, 
  and \textsf{kernel}, which computes the total energy.
   
\subsection{Case study: implementing an embedding model}
\begin{figure*}
	\centering
	\includegraphics{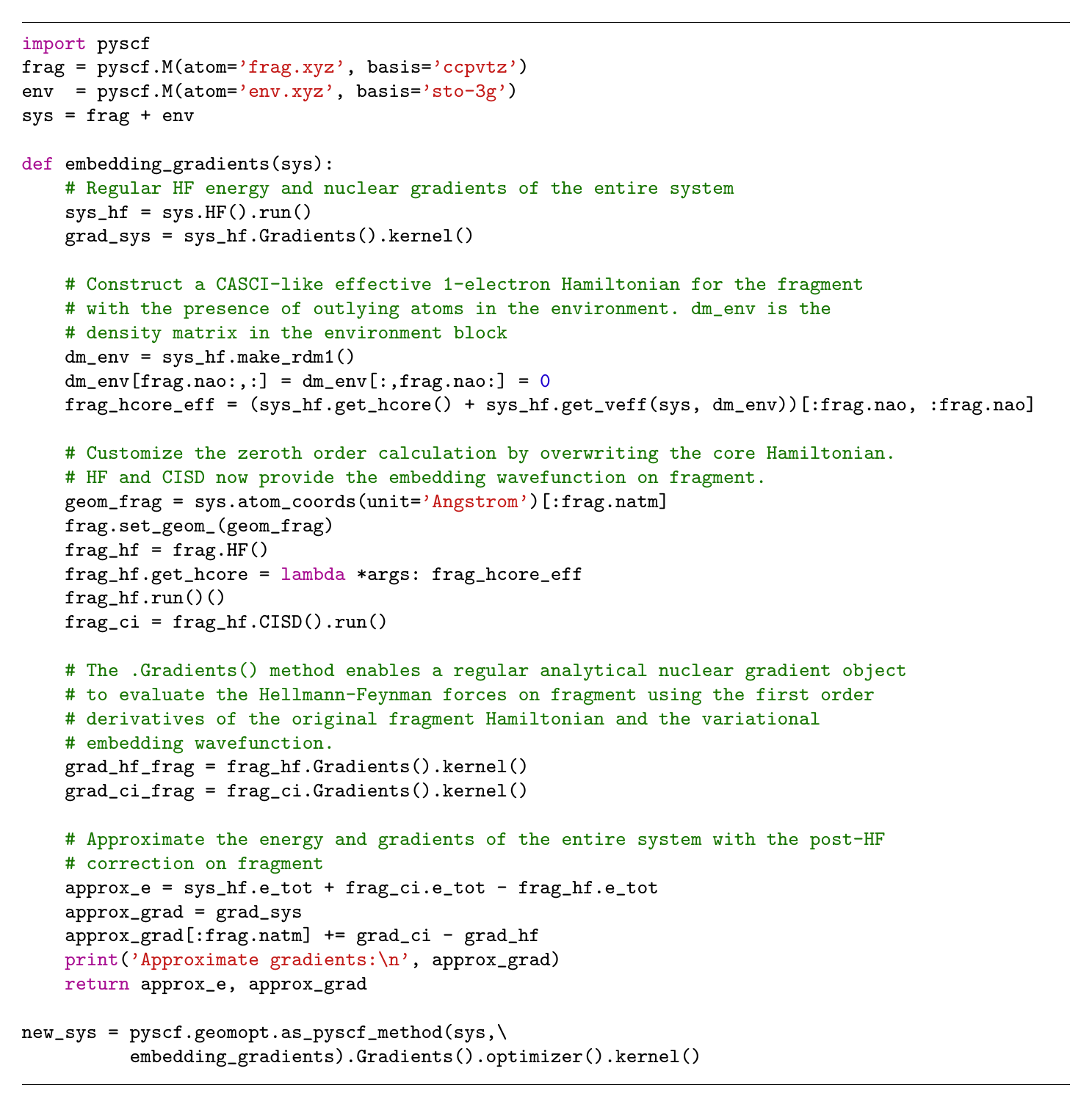}
	\caption{An advanced example that implements geometry optimization based on a WFT-in-HF embedding model using standard \pyscf APIs.}
	\label{code:geomopt}
\end{figure*}

As a more advanced example of customization using \pyscf, we now
illustrate how a simple script with standard APIs enables \pyscf to carry out geometry optimization
for a wavefunction in Hartree-Fock (WFT-in-HF) embedding model, shown in Figure \ref{code:geomopt}
with a CISD solver.
Given the Hamiltonian of a system, expressed in terms of the Hamiltonians of a fragment and its environment
\begin{gather*}
  H_\text{sys} = H_\text{frag} + H_\text{env} + V_\text{ee,frag-env}, \\
  H_\text{frag} = h_\text{core,frag} + V_\text{ee,frag}, \\
  H_\text{env} = h_\text{core,env} + V_\text{ee,env},
\end{gather*}
we define an embedding Hamiltonian for the fragment
in the presence of the atoms in the environment as
\begin{gather*}
  H_\text{emb} = h_\text{eff,frag} + V_\text{ee,frag}, \\
  h_\text{eff,frag} = h_\text{core,frag} + (h_\text{core,env} + V_\text{eff}[\rho_\text{env}]),\\
  V_\text{eff}[\rho_\text{env}]
  = \int V_\text{ee,env} \rho_\text{env}(\mathbf{r}) d \mathbf{r}
  + \int V_\text{ee,frag-env} \rho_\text{env}(\mathbf{r}) d \mathbf{r}
\end{gather*}
Geometry optimization can then be carried out with the approximate
nuclear gradients of the embedding problem
\begin{widetext}
\begin{align*}
\text{Gradients}
&=      \langle\Psi_\text{CI}|\frac{\partial H_{sys}}{\partial X}|\Psi_\text{CI}\rangle \\
&\approx \langle\Psi_\text{CI}|\frac{\partial H_\text{frag}}{\partial X}|\Psi_\text{CI} \rangle
+        \langle\Psi_\text{HF}|\frac{\partial (H_\text{env} + V_\text{ee,frag-env})}{\partial X}|\Psi_\text{HF}\rangle \\
&=       \langle\Psi_\text{CI}|\frac{\partial H_\text{frag}}{\partial X}|\Psi_\text{CI} \rangle
-        \langle\Psi_\text{HF}|\frac{\partial H_\text{frag}}{\partial X}|\Psi_\text{HF}\rangle
+        \langle\Psi_\text{HF}|\frac{\partial H_\text{sys} }{\partial X}|\Psi_\text{HF} \rangle \\
&\approx \langle\Psi_\text{frag,CI}|\frac{\partial H_\text{frag}}{\partial X}|\Psi_\text{frag,CI}\rangle
-        \langle\Psi_\text{frag,HF}|\frac{\partial H_\text{frag}}{\partial X}|\Psi_\text{frag,HF}\rangle
+        \langle\Psi_\text{HF}|\frac{\partial H_\text{sys}}{\partial X}|\Psi_\text{HF} \rangle,
\end{align*}
\end{widetext}
where the fragment wavefunction $\Psi_\text{frag,HF}$ and $\Psi_\text{frag,CI}$ are obtained
from the embedding Hamiltonian $H_\text{emb}$. The code snippet 
in Figure \ref{code:geomopt} demonstrates
the kind of rapid prototyping that can be carried out using \pyscf APIs.
In particular, this demonstration combines the APIs for ab initio energy evaluation, analytical nuclear
gradient computation, computing the HF potential for an arbitrary density matrix, Hamiltonian
customization, and customizing the nuclear gradient solver in a geometry optimization.

\section{Summary of existing methods and recent additions}\label{sec:methods}

In this section we briefly summarize major current capabilities of the \pyscf package.
These capabilities are listed in Table~\ref{tb:feature} and details 
are presented in the following subsections.

\begin{table*}
	\centering
	\begin{threeparttable}
	\caption{
		Major features of \pyscf as of version~1.7.1.
	}
	\label{tb:feature}
	\begin{tabular*}{1.0\textwidth}{@{\extracolsep{\fill}}lllp{10cm}}
                \hline\hline
Methods            & Molecules & Solids  & Comments \\
\hline
HF                 & Yes      & Yes     & $\sim$ 10000 AOs\tnote{a} \\
MP2                & Yes      & Yes     & $\sim$ 1500 MOs\tnote{a} \\
DFT                & Yes      & Yes     & $\sim$ 10000 AOs\tnote{a} \\
TDDFT/TDHF/TDA/CIS & Yes      & Yes     &$\sim$ 10000 AOs\tnote{a} \\
$\rm G_0W_0$       & Yes      & Yes     & $\sim$ 1500 MOs\tnote{a} \\
CISD               & Yes      & Yes\tnote{b} &$\sim$ 1500 MOs\tnote{a} \\
FCI                & Yes      & Yes\tnote{b}  &$\sim$ (18e, 18o)\tnote{a} \\
IP/EA-ADC(2)         & Yes      & No      & $\sim$ 500 MOs\tnote{a,c} \\
IP/EA-ADC(2)-X     & Yes      & No      & $\sim$ 500 MOs\tnote{a,c} \\
IP/EA-ADC(3)         & Yes      & No      & $\sim$ 500 MOs\tnote{a,c} \\
CCSD               & Yes      & Yes     & $\sim$ 1500 MOs\tnote{a} \\
CCSD(T)            & Yes      & Yes     & $\sim$ 1500 MOs\tnote{a} \\
IP/EA/EE-EOM-CCSD\tnote{d}            & Yes      & Yes     & $\sim$ 1500 MOs\tnote{a} \\
MCSCF              & Yes      & Yes\tnote{b}  & $\sim$ 3000 AOs,\tnote{a} \ 30--50 active orbitals\tnote{e} \\
MRPT               & Yes      & Yes\tnote{b}  & $\sim$ 1500 MOs,\tnote{a} \ 30--50 active orbitals\tnote{e} \\
QM/MM              & Yes      & No      & \\
Semiempirical      & Yes      & No      & MINDO3 \\
Relativity         & Yes      & No      & ECP and scalar-relativistic corrections for all methods. 2-component methods for HF, DFT, DMRG and SHCI. 4-component methods for HF and DFT.\\
Gradients          & Yes      & No      & HF, MP2, DFT, TDDFT, CISD, CCSD, CCSD(T), MCSCF and MINDO3 \\
Hessian            & Yes      & No      & HF and DFT \\
Orbital Localizer  & Yes      & Yes     & NAO, meta-L\"{o}wdin, IAO/IBO, VVO/LIVVO, Foster-Boys, Edmiston–Ruedenberg, Pipek–Mezey and Maximally-localized Wannier functions \\
Properties         & Yes      & Yes\tnote{f}      & EFGs, M\"{o}ssbauer spectroscopy, NMR, magnetizability, and polarizability, {\em etc.}\\
Solvation          & Yes      & No      & ddCOSMO, ddPCM, and polarizable embedding \\
AO, MO integrals & Yes & Yes   & 1-electron and 2-electron integrals \\
Density fitting    & Yes      & Yes     & HF, DFT, MP2 and CCSD \\
Symmetry           & Yes      & No\red{\tnote{g}}     & $D_{2h}$ and subgroups for \red{molecular} HF, MCSCF, and FCI\\
		\hline\hline
	\end{tabular*}
	\begin{tablenotes}
		\item[a] An estimate based on a single SMP node with 128 GB memory without density fitting;
		\item[b] $\Gamma$-point only; 
		\item[c] In-core implementation limited by storing two-electron integrals in memory;
		\item[d] Perturbative corrections to IP and EA via IP-EOM-CCSD* and EA-EOM-CCSD* are available for both molecules and crystals;
		\item[e] Using an external DMRG, SHCI, or FCIQMC program \red{(as listed in Section \ref{sec:orbital_opt})} as the active space solver;
		\item[f] EFGs and M\"{o}ssbauer spectra only;
		\red{\item[g] Experimental support for point-group and time-reversal symmetries in crystals at the SCF and MP2 levels.}
	\end{tablenotes}
    \end{threeparttable}
\end{table*}

\subsection{Hartree-Fock and density functional theory methods}

\begin{figure*}
\includegraphics{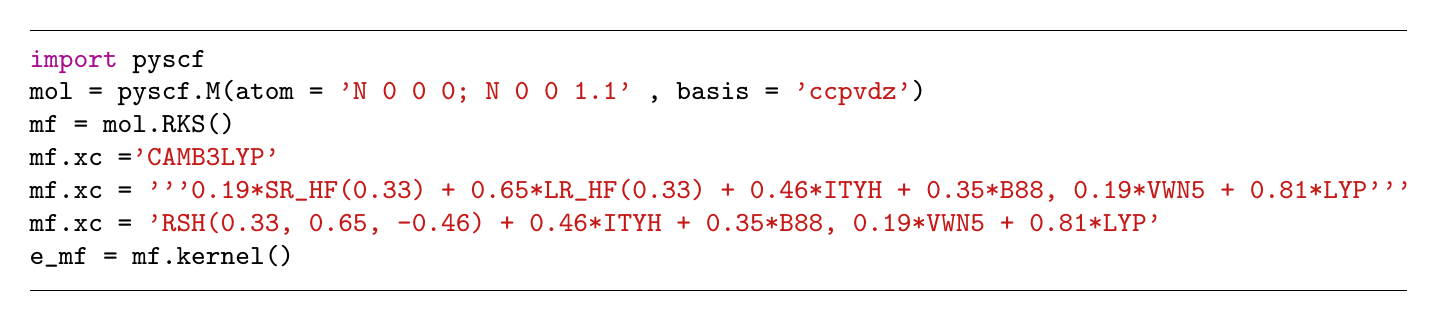}
\caption{An example of two customized RSH functionals that are equivalent to the CAM-B3LYP functional.}
\label{code:rsh}
\end{figure*}

The starting point for many  electronic structure simulations is a 
self-consistent field (SCF) calculation. 
\pyscf implements Hartree-Fock (HF) and density functional theory (DFT) 
with a variety of Slater determinant references, 
including restricted closed-shell, restricted open-shell, unrestricted,
and generalized (noncollinear spin) references,\red{\cite{Seeger1977,VanWullen2002}}
for both molecular and crystalline ($k$-point) calculations.
Through an interface to the \textsc{Libxc}\cite{Lehtola2018} and \textsc{XCFun}\cite{Ekstrom2010} libraries, \pyscf also supports a wide range of 
predefined exchange-correlation (XC) functionals, including the local density approximation (LDA),
generalized gradient approximations (GGA), hybrids, meta-GGAs, nonlocal correlation functionals (VV10\cite{Vydrov2010}) and range-separated hybrid (RSH) functionals. 
In addition to  predefined XC functionals, the user can also create customized functionals in a DFT calculation, as shown in Figure \ref{code:rsh}.
  
Because \pyscf uses a Gaussian AO representation, the SCF computation 
is usually dominated by Gaussian integral evaluation. Through the efficient Gaussian integral engine \textsc{Libcint},\cite{Sun2015}
the molecular SCF module can be used with more than 10,000 basis functions on a symmetric multiprocessing (SMP) machine, without resorting to any
integral approximations such as screening. Further speed-up can be achieved through Gaussian density fitting, and 
the pseudo-spectral approach (SGX) is implemented to speed up the evaluation of exchange in large systems.\cite{Friesner1985,Neese2009,Izsak2011}

\begin{figure}
	\includegraphics{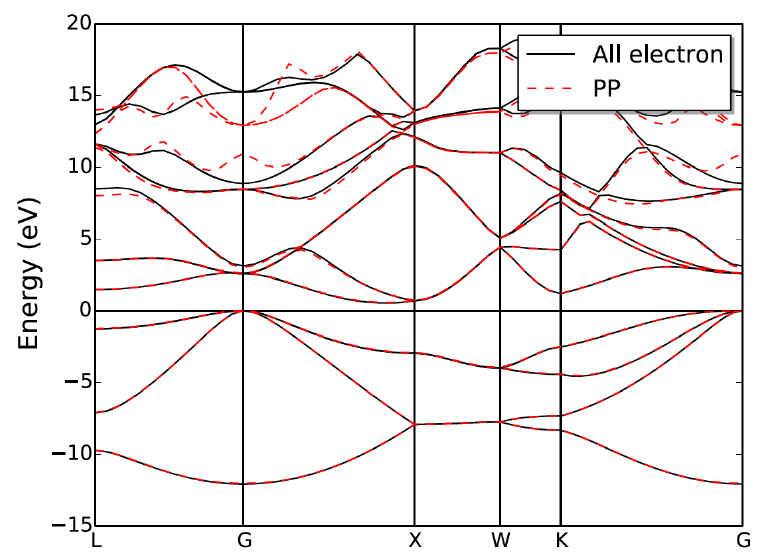}
	\caption{All-electron and pseudopotential LDA band structures of the Si crystal.
		Reprinted from Ref.~\onlinecite{Sun2017}, with the permission of AIP Publishing.}
	\label{fig:si_band}
\end{figure}

In crystalline systems, HF and DFT calculations can be carried out either at a single point in the Brillouin zone or with a $k$-point mesh.
The cost of the crystalline SCF calculation depends on the nature of the crystalline Gaussian basis and the associated density
fitting. \pyscf supports Goedecker-Teter-Hutter (GTH) pseudopotentials \cite{Goedecker1996} 
which can be used with the associated basis sets (developed by the CP2K group).\cite{VandeVondele2005,Hutter2014} Pseudopotential DFT calculations are typically most efficiently done using plane-wave density fitting (FFTDF). Alternatively,
all-electron calculations can be performed with standard basis sets, and the presence of sharp densities means that
Gaussian density fitting performs better. Gaussian density fitting is also the algorithm
of choice for calculations with HF exchange.
Figure \ref{fig:si_band} shows an example of the silicon band structures computed using a GTH-LDA pseudopotential with FFTDF, and in an all-electron calculation using GDF.

\subsection{Many-body methods}


Starting from a SCF HF or DFT wavefunction, various many-body methods are available in \pyscf,
including M{\o}ller-Plesset second-order perturbation theory (MP2), 
multi-reference perturbation theory (MRPT),\cite{Angeli2001,Guo2016b} 
configuration interaction (CI),\cite{Langhoff1974,Pople1977,Knowles1984,Olsen1990} 
coupled cluster (CC),\cite{Sekino1984,Scheiner1987,Scuseria1988,Raghavachari1989,Salter1989,Scuseria1991,Nooijen1995,Koch1996,Musial2003,Krylov2006a}
multi-configuration self-consistent field (MCSCF),\cite{Werner1985,Jensen1987}
algebraic diagrammatic construction (ADC)\cite{Schirmer1982,Schirmer1983,Schirmer2004,Dreuw2015,Banerjee2019} 
and $\rm G_0W_0$\cite{Hedin1965,Aryasetiawan1998,Ren2012,Wilhelm2016} methods.
The majority of these capabilities are available for both molecules and crystalline materials.

\subsubsection{Molecular implementations}

The \pyscf CI module implements solvers for configuration interaction with single and 
double excitations (CISD), and a general full configuration interaction (FCI) solver that
 can treat fermion, boson and coupled fermion-boson Hamiltonians.
The FCI solver is heavily optimized for its multithreaded performance
and can efficiently handle active spaces with up to 18 electrons in 18 orbitals.

\begin{figure*}
	\includegraphics{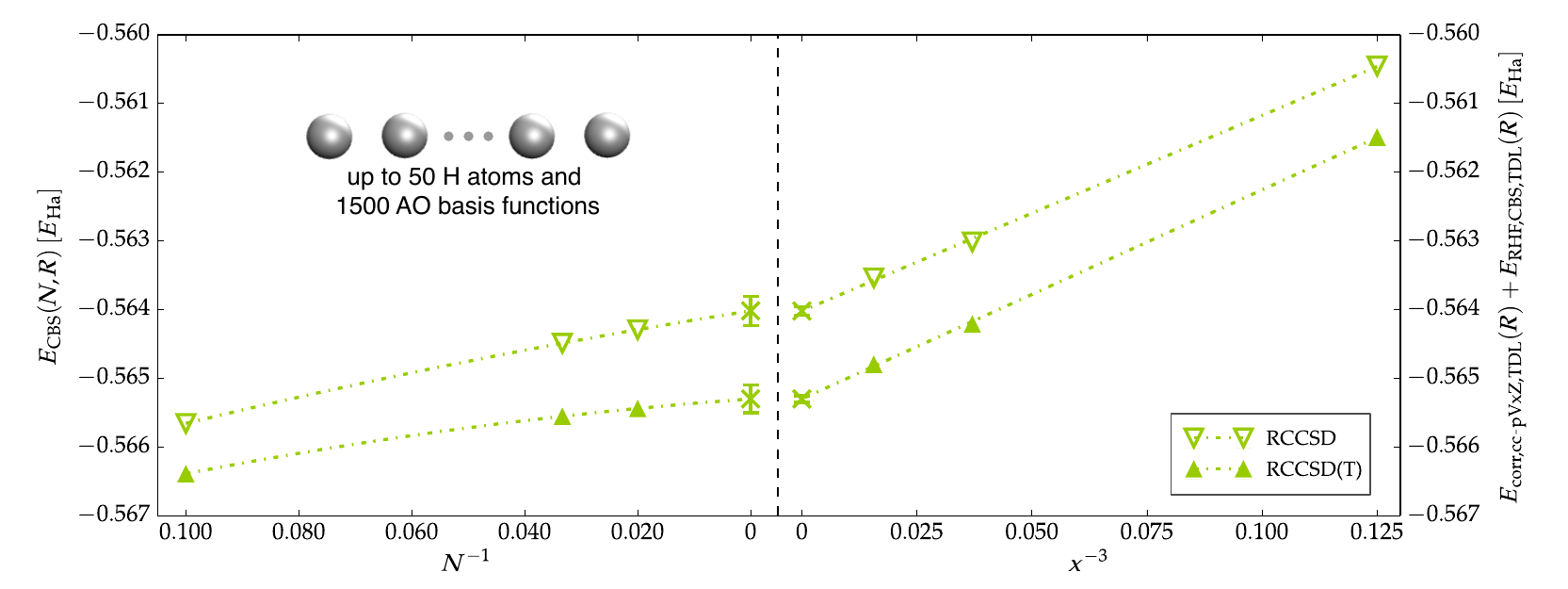}
	\caption{Energies of a hydrogen chain computed at the restricted CCSD and CCSD(T) levels
		 extrapolated to the complete basis set (CBS) and thermodynamic limits. 
		 The left-hand panel shows extrapolation of $E_{\rm CBS}(N)$ versus $1/N$, 
		 where $N$ is the number of atoms;
		 while the right-hand panel shows extrapolation of $E_{\rm cc-pVxZ}(N\rightarrow\infty)$ versus $1/x^3$ with x equal to 2, 3 and 4 corresponding to 
		 double-, triple- and quadruple-zeta basis, respectively.
		 Adapted from Ref.~\onlinecite{Motta2017}.
	}
	\label{fig:h50}
\end{figure*}

The CC module implements coupled cluster theory with single and double excitations (CCSD)\cite{Scuseria1988,Koch1996}
and with the perturbative triples correction [CCSD(T)].\cite{Raghavachari1989} 
$\Lambda$-equation solvers are implemented to compute 
one- and two-particle density matrices, as well as 
the analytic nuclear gradients for the CCSD and CCSD(T) methods.\cite{Scheiner1987,Salter1989,Scuseria1991}
\pyscf also implements various flavours of equation-of-motion CCSD
to compute electron affinities (EA), ionization potentials (IP), neutral excitation energies (EE), and spin-flip excitation energies (SF).\cite{Sekino1984,Nooijen1995,Musial2003,Krylov2006a} Experimental support for
beyond doubles corrections to IP and EA via IP-EOM-CCSD*\red{\cite{Stanton1996,Saeh1999}} and EA-EOM-CCSD* is also available. 
For very large basis sets, \pyscf provides an efficient AO-driven pathway which allows calculations with more than 1500 basis functions.
An example of this is shown in Figure \ref{fig:h50}, where the largest CCSD(T) calculation 
contains 50 electrons and 1500 basis functions.\cite{Motta2017}

Second- and third-order algebraic diagrammatic construction (ADC) methods are also available 
in \pyscf for the calculation of molecular electron affinities
and ionization potentials\cite{Schirmer1982,Schirmer1983,Schirmer2004,Dreuw2015,Banerjee2019}
[EA/IP-ADC(n), n = 2, 3]. These have a lower cost than
EA/IP-EOM-CCSD. The advantage of the ADC methods over EOM-CCSD is that their 
amplitude equations can be solved in one iteration and the eigenvalue problem is Hermitian, 
which lowers the cost of computing the EA/IP energies and transition intensities. 

\begin{figure}
	\includegraphics{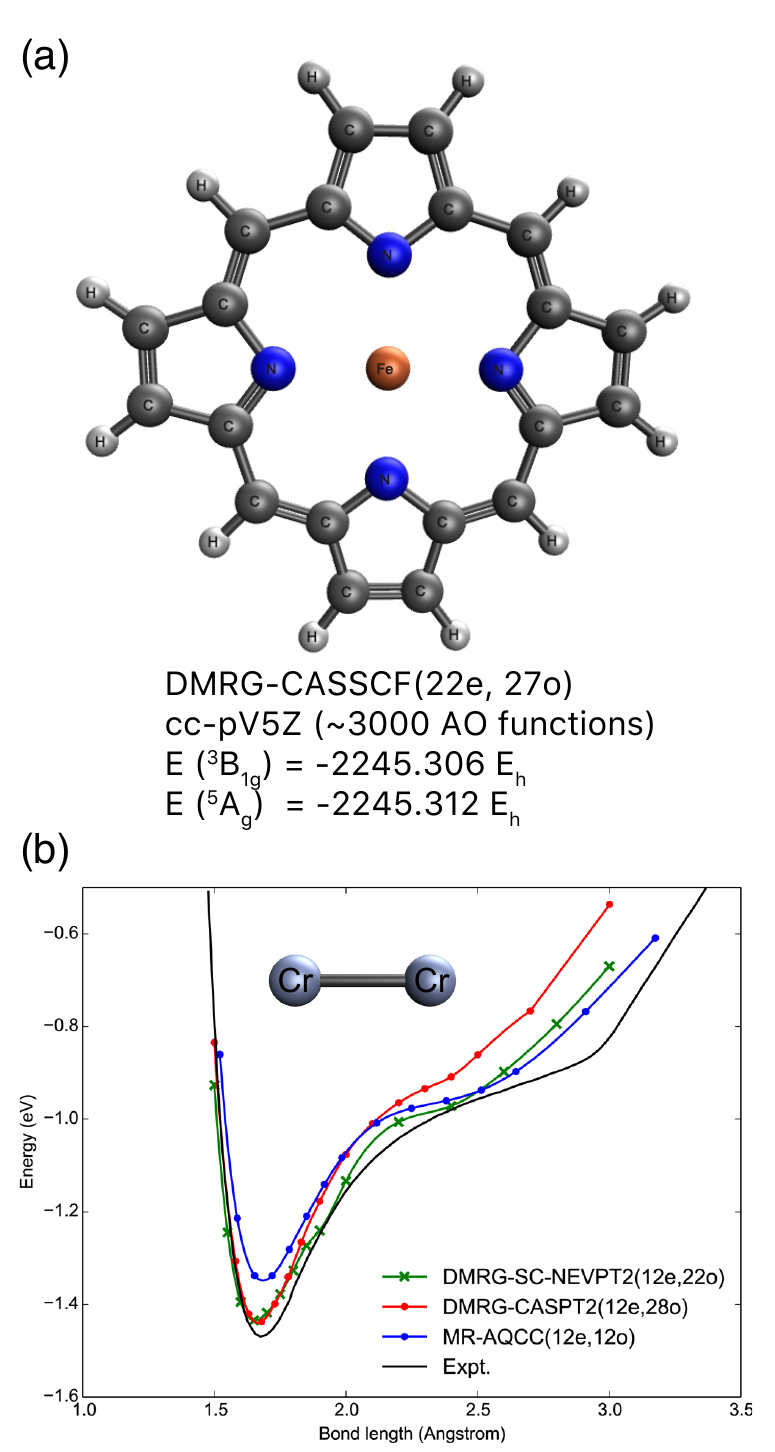}
	\caption{(a) Ground-state energy calculations for Fe(II)-porphine at the DMRG-CASSCF/cc-pV5Z 
		level with an active space of 22 electrons in 27 orbitals.\cite{Sun2017a}
		(b) Potential energy curve for $\rm Cr_2$ at the DMRG-SC-NEVPT2 (12e, 22o) level, compared
		to the results from other methods. 
		Adapted with permission from Ref.~\onlinecite{Guo2016b}. Copyright (2016) American Chemical Society.
	}
	\label{fig:FeP}
\end{figure}

The MCSCF module provides complete active space configuration interaction (CASCI) and
complete active space self-consistent field (CASSCF)\cite{Werner1985,Jensen1987} methods
for multi-reference problems.
As discussed in section~\ref{sec:orbital_opt}, the module also provides a general second-order orbital optimizer\cite{Sun2017a} that
can optimize the orbitals of external methods, with native interfaces for the orbital optimization of 
density matrix renormalization group (DMRG),\cite{Sharma2012,Wouters2014}
full configuration interaction quantum Monte Carlo (FCIQMC),\cite{Booth2014,Thomas2015} 
 and selected configuration interaction wavefunctions.\cite{Sharma2017,Smith2017}
Starting from a CASCI or CASSCF wavefunction, \pyscf also implements
the strongly-contracted second-order $n$-electron valence perturbation theory\cite{Angeli2001,Guo2016b} (SC-NEVPT2)
in the MRPT module to include additional dynamic correlation. Together with external active-space solvers this enables one to treat relatively large active spaces for such calculations, as illustrated in Figure \ref{fig:FeP}.

\subsubsection{Crystalline implementations}
\begin{figure*}
\includegraphics{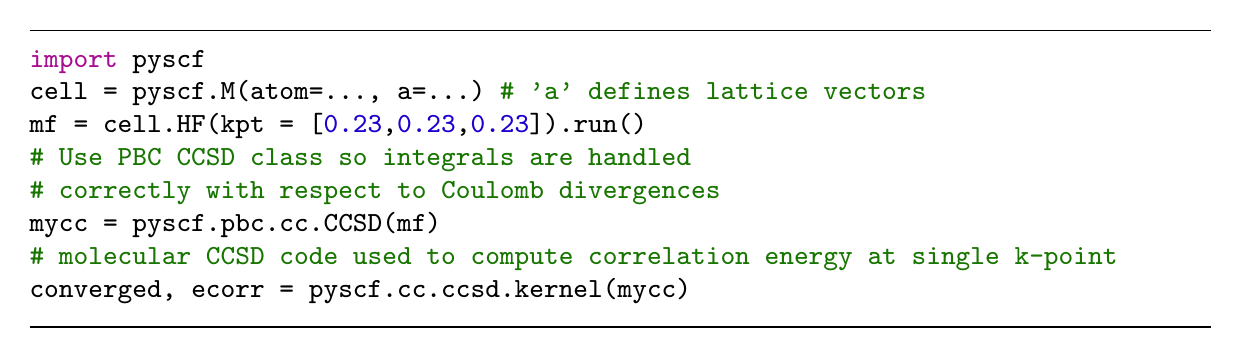}
\caption{Illustration of using the molecular code to compute an energy in crystal at a single $k$-point.}
\label{code:singlek}
\end{figure*}

\begin{figure}
	\includegraphics{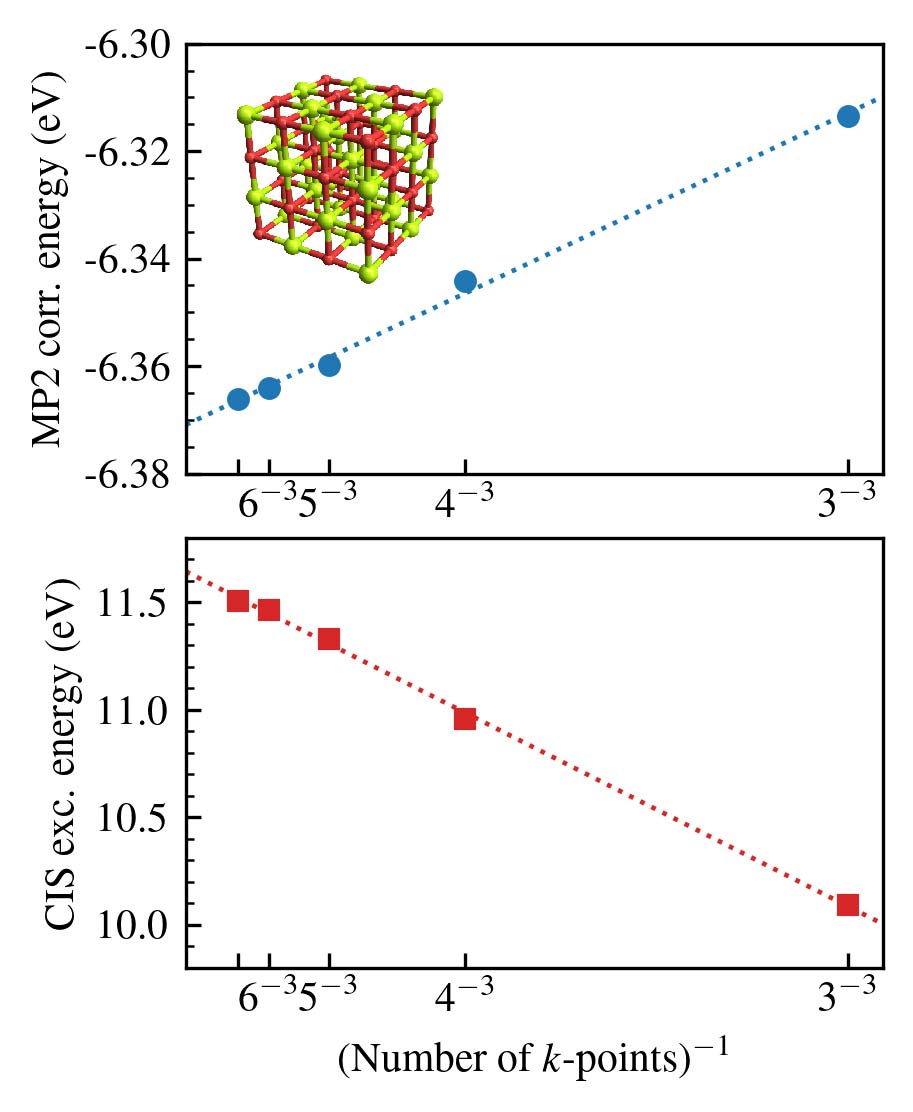}
        \caption{Periodic MP2 correlation energy per unit cell (top) and CIS
excitation energy (bottom) as a function of the number of $k$-points sampled in
the Brillouin zone for the MgO crystal.}
	\label{fig:mp2_cis}
\end{figure}

\begin{figure*}
	\includegraphics{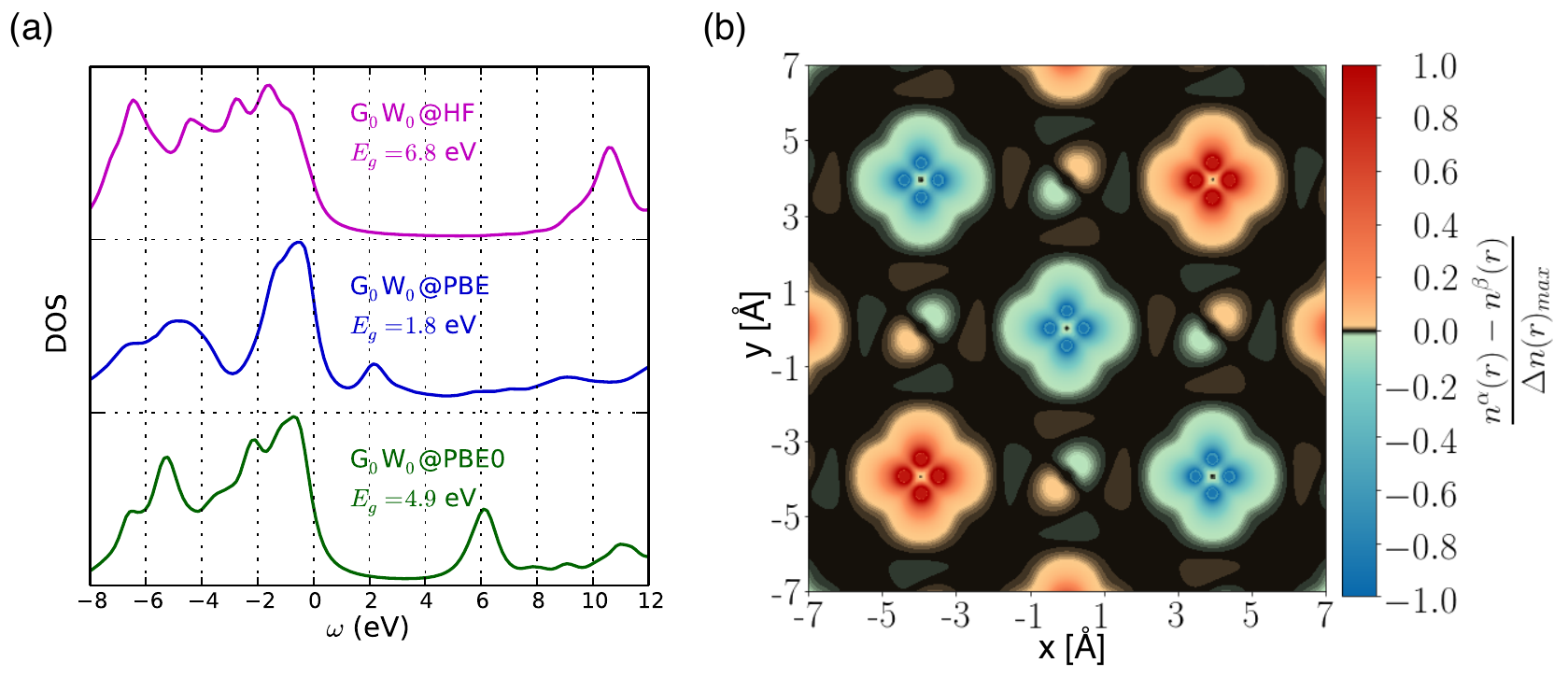}
	\caption{Electronic structure calculations for antiferromagnetic NiO.
		(a) Density of states and band gaps computed by $\rm G_0W_0$.
		(b) Normalized spin density on the (100) surface by CCSD (the Ni atom is
		located at the center). Adapted from Ref.~\onlinecite{Gao2019}.}
	\label{fig:gw}
\end{figure*}

As discussed in section~\ref{sec:crystalline}, the \pyscf
implementations of many-body methods for crystalline systems closely parallel their molecular implementations.
In fact, all molecular modules can be used to carry out
calculations in solids at the $\Gamma$-point and many modules (those supporting complex integrals)
can be used at any other single $k$-point.
Such single $k$-point calculations only require
the appropriate periodic integrals to be supplied to the many-body solver (Figure \ref{code:singlek}).
For those modules that support complex integrals, twist averaging can then be performed to sample the Brillouin zone.
To use savings from $k$-point symmetries, an additional summation over momentum
conserving $k$-point contributions needs to be explicitly implemented. Such
implementations are provided for MP2, CCSD, CCSD(T),
IP/EA-EOM-CCSD\cite{McClain2017} and 
EE-EOM-CCSD,\cite{Wang2020} and $\rm G_0W_0$.
For example, Figure~\ref{fig:mp2_cis} shows the MP2 correlation energy and
the CIS excitation energy of MgO, calculated using periodic
density-fitted implementations; the largest system shown, with a 
$7\times 7\times 7$ $k$-point mesh, correlates 5,488 valence electrons in
9,261 orbitals.
Furthermore, Figure \ref{fig:gw} shows some examples of periodic correlated calculations on NiO carried out
using the $\rm G_0W_0$ and CCSD methods.

\subsection{Efficiency}
In Table~\ref{tb:feature}, we provide rough estimates of the sizes of problems
that can be tackled using \pyscf for each electronic structure method.
Figs.~\ref{fig:h50}, \ref{fig:FeP}, \ref{fig:mp2_cis}, and \ref{fig:gw} illustrate
some real-world examples of calculations performed using \pyscf. 
Note that the size of system that can be treated is a function of the computational resources available; the
estimates given above assume relatively standard and modest computational resources, e.g. a node
of a cluster, or a few dozen cores.
For more details of the runtime environment and program settings for
similar performance benchmarks, we refer readers to the benchmark page of the PySCF website
\url{www.pyscf.org}.
The implementation and performance of \pyscf on massively parallel architectures is discussed in
section~\ref{sec:mpi}.


For molecular calculations using mean-field methods, \pyscf can treat systems with more than 10,000 AO basis functions without difficulty.
Fig.~\ref{fig:mpi} shows the time of building the HF Fock matrix for a large water cluster with more than 12,000 basis functions. 
\blue{With the integral screening threshold set to $10^{-13}$ a.u.,} it takes only around 7 hours on one computer node with 32 CPU cores.
Applying MPI parallelization further reduces the Fock-build time (see section~\ref{sec:mpi}).
For periodic boundary calculations at the DFT level using pure XC functionals, 
even larger systems can be treated using pseudopotentials and a multi-grid implementation.
Table~\ref{tb:dft_time} presents an example of such a calculation,
where for the largest system considered (\text{[H$_2$O]$_{512}$} with more
than 25,000 basis functions), the Fock-build time is about an hour or less on a single node.

To demonstrate the efficiency of the many-body method implementations, 
in Tables~\ref{tb:ccsd_time} and \ref{tb:fci_time}
we show timing data of exemplary CCSD and FCI calculations. 
It is clear that systems with more than 1,500 basis functions 
can be easily treated at the CCSD level and that the FCI implementation in \pyscf is 
very efficient. In a similar way, the estimated performance for other many-body methods implemented in \pyscf is
listed in Table~\ref{tb:feature}.

\begin{table}
	\centering
	\begin{threeparttable}
		\caption{
			Wall time (in seconds) for building the Fock matrix in a supercell DFT calculation
			of water clusters with the GTH-TZV2P\cite{VandeVondele2005,Hutter2014} basis set, using the SVWN\cite{Vosko1980} and the 
			PBE\cite{Perdew1996} XC functionals and corresponding pseudopotenials, respectively.
            \blue{Integral screening and lattice summation cutoff were controlled by an overall threshold of $10^{-6}$ a.u. for Fock
            matrix elements.}
            The calculations were performed on one computer node with 32 
			\blue{Intel Xeon Broadwell (E5-2697v4) processors}.
		}
		\label{tb:dft_time}
		\begin{tabular*}{.45\textwidth}{@{\extracolsep{\fill}}lccc}
			\hline\hline
			System            & $N_{\rm AO}$\tnote{a} & SVWN & PBE \\
			\hline
			\text{[H$_2$O]$_{32}$}                & 1,280     & 8        &23 \\
			\text{[H$_2$O]$_{64}$}                & 2,560     & 20      &56 \\
			\text{[H$_2$O]$_{128}$}              & 5,120     & 74      &253\\
			\text{[H$_2$O]$_{256}$}              & 12,800   & 276    &1201\\
			\text{[H$_2$O]$_{512}$} 			 & 25,600   & 1279  &4823\\
			\hline\hline
		\end{tabular*}
	\begin{tablenotes}
		\item[a] Number of AO basis functions.
	\end{tablenotes}
	\end{threeparttable}
\end{table}

\begin{table}
	\centering
	\begin{threeparttable}
		\caption{
			Wall time (in seconds) for the first CCSD iteration in \blue{AO-driven} CCSD calculations on hydrogen chains.
            \blue{The threshold of integral screening was set to $10^{-13}$ a.u..
            For these hydrogen chain molecules, CCSD takes around 10 iterations to converge.
			The calculations were performed on one computer node with 28 
			Intel Xeon Broadwell (E5-2697v4) processors}.
		}
		\label{tb:ccsd_time}
		\begin{tabular*}{.45\textwidth}{@{\extracolsep{\fill}}lccc}
			\hline\hline
			System/basis set      & $N_{\rm occ}$\tnote{a} &$N_{\rm virt}$\tnote{b} & time \\
			\hline
			H$_{30}$/cc-pVQZ  &15   &884               &621 \\
			H$_{30}$/cc-pV5Z  &15  &1631            &6887 \\
			H$_{50}$/cc-pVQZ  &25   &1472             &8355 \\
			\hline\hline
		\end{tabular*}
	\begin{tablenotes}
		\item[a] Number of active occupied orbitals;
		\item[b] Number of active virtual orbitals.
	\end{tablenotes}
	\end{threeparttable}
\end{table}

\begin{table}
	\centering
	\begin{threeparttable}
		\caption{
			Wall time (in seconds) for one FCI iteration for different active-space sizes.
			The calculations were performed on one computer node with 32 
			\blue{Intel Xeon Broadwell (E5-2697v4) processors}.
		}
		\label{tb:fci_time}
		\begin{tabular*}{.45\textwidth}{@{\extracolsep{\fill}}lc}
			\hline\hline
			Active space      &time \\
			\hline
			(12e, 12o) &0.1 \\
			(14e, 14o) &0.7 \\
			(16e, 16o) &8 \\
			(18e, 18o) &156 \\
			\hline\hline
		\end{tabular*}
	\end{threeparttable}
\end{table}

\subsection{Properties}
At the mean-field level, the current \pyscf program 
can compute various nonrelativistic and four-component relativistic molecular properties. 
These include NMR shielding and spin-spin coupling tensors,\cite{Visscher1999,Helgaker1999,Enevoldsen2000,Sychrovsky2000,Helgaker2000,Cheng2009}
electronic g-tensors,\cite{Schreckenbach1997,Neese2001,Rinkevicius2003,Hrobarik2011}
nuclear spin-rotation constants and rotational g-tensors,\cite{Sauer1992,Gauss1996}
hyperfine coupling (HFC) tensors,\cite{Neese2003,Arbuznikov2004} 
electron spin-rotation (ESR) tensors,\cite{Curl1965,Tarczay2010} 
magnetizability tensors,\cite{Sauer1992,Keith1996,Cammi1998}
zero-field splitting (ZFS) tensors,\cite{Pederson1999,Neese2007,Schmitt2011} 
as well as static and dynamic polarizability and hyper-polarizability tensors.
The contributions from spin-orbit coupling and spin-spin coupling can also be
calculated and included in the g-tensors, HFC tensors, ZFS tensors, and ESR
tensors. In magnetic property calculations, approximate gauge-origin invariance is ensured for NMR shielding,
g-tensors, and magnetizability tensors via the use of gauge including atomic
orbitals.\red{\cite{London1937,Ditchfield1974,Sauer1992,Keith1996,Cammi1998}}

Electric field gradients (EFGs) and M\"{o}ssbauer parameters\cite{Petrilli1998,Adiga2007,Autschbach2010} can be computed 
using either the mean-field electron density, or the correlated density obtained from
non-relativistic Hamiltonians, spin-free \red{exact-two-component (X2C)} relativistic Hamiltonians\red{\cite{Dyall1997,Kutzelnigg2005,Liu2006,Ilias2007,Cheng2011}} or
four-component methods, in both molecules and crystals.

Finally, analytic nuclear gradients for the molecular ground state are
available at the mean-field level and for many of  the electron correlation methods such as
MP2, CCSD, CISD, CASCI and CASSCF (see Table \ref{tb:feature}). The CASCI gradient implementation supports the use of external solvers, 
such as DMRG, and provides gradients for such methods.
\pyscf also implements the analytical gradients of \red{time-dependent density functional theory 
(TDDFT) with or without the Tamm-Dancoff approximation (TDA)}
for excited state geometry optimization.
The spin-free X2C relativistic Hamiltonian, frozen core approximations, solvent
effects, and molecular mechanics (MM) environments can be combined with any of the nuclear gradient methods.
Vibrational frequency and thermochemical analysis can also be performed, using the
analytical Hessians from mean-field level calculations, or numerical Hessians
of methods based on numerical differentiation of analytical gradients.


\subsection{Orbital localization}
\pyscf provides two kinds of orbital localization in the LO module.
The first kind localizes orbitals based on the atomic character of the basis functions,
and 
can generate intrinsic atomic orbitals (IAOs),\cite{Knizia2013a}
natural atomic orbitals (NAOs),\cite{Reed1985} and
meta-L\"{o}wdin orbitals.\cite{Sun2014}
These AO-based local orbitals can be used to carry out reliable population analysis
in arbitrary basis sets.

The second kind optimizes a cost function to produce localized orbitals.
\pyscf implements Boys localization,\cite{Foster1960} Edmiston-Ruedenberg localization,\cite{Edmiston1965} 
and Pipek–Mezey localization.\cite{Pipek1989}
Starting from the IAOs, one can also use orbital localization based on the Pipek-Mezey procedure
to construct the intrinsic bond orbitals (IBOs).\cite{Knizia2013a}
A similar method can also be used to construct localized intrinsic valence virtual orbitals
that can be used to assign core-excited states.\cite{Derricotte2017}
The optimization in these localization routines takes advantage of the 
second order coiterative augmented Hessian (CIAH) algorithm\cite{Sun2016a} for rapid convergence.

\begin{figure}
	\begin{center}
		\includegraphics{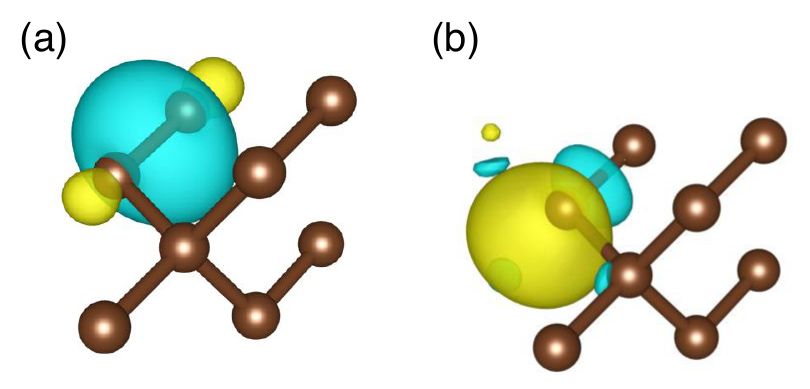}
	\end{center}
	\caption{(a) IBOs for diamond at the $\Gamma$-point (showing one $\sigma$ bond);
	         (b) MLWFs for diamond computed within the valence IAO subspace (showing one $\rm sp^3$ orbital).}
	\label{fig:ibo}
\end{figure}

For crystalline calculations with $k$-point sampling, 
\pyscf also provides maximally-localised Wannier functions (MLWFs) 
via a native interface to the \textsc{Wannier90} program.\cite{Pizzi2020}
Different types of orbitals are available as  initial guesses for the MLWFs, 
including the atomic orbitals provided by \textsc{Wannier90}, meta-L\"{o}wdin orbitals,\cite{Sun2014}
and localized orbitals from the selected columns of density matrix (SCDM) method.\cite{Damle2015,Damle2017}
Figure \ref{fig:ibo} illustrates the IBOs and MLWFs of diamond computed by \pyscf.

\subsection{QM/MM and solvent}

\begin{figure}
	\includegraphics{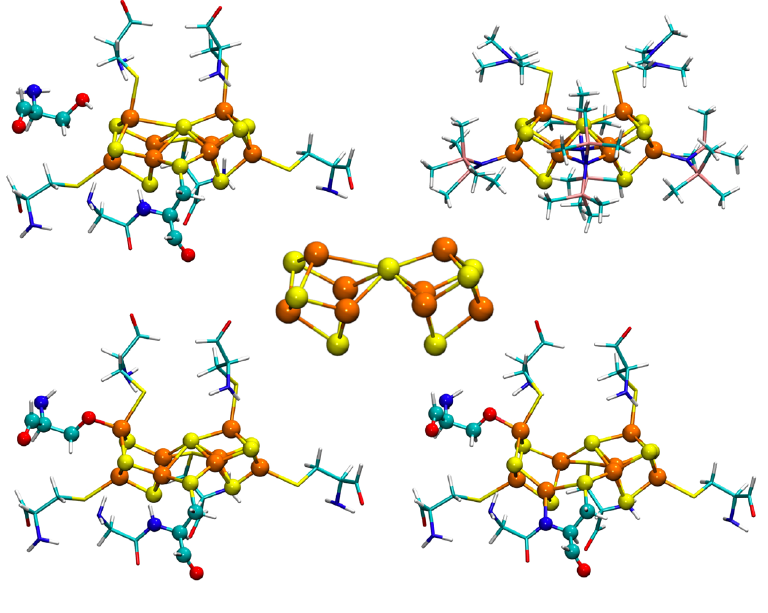}
	\caption{Illustration of P-cluster \red{(the [Fe$_8$S$_7$] cluster of nitrogenase)} calculations where the COSMO solvation model was used to mimic the protein environment of nitrogenase
          beyond the first coordination sphere. 
          \red{Fe, orange; S, yellow; C, cyan; O, red; N, blue; H, white; Si, pink.}
          Adapted from Ref.~\onlinecite{Li2019}. 
        }
	\label{fig:solvent}
\end{figure}

\pyscf incorporates two continuum solvation models, namely, 
the conductor-like screening model\cite{Klamt1993} (COSMO) and 
the polarizable continuum model using the integral equation formalism\cite{Cances1997,Mennucci1997} (IEF-PCM).
Both of them are implemented  efficiently via a 
domain decomposition (dd) approach,\cite{Cances2013,Lipparini2013,Lipparini2014,Stamm2016,Lipparini2016} 
and are compatible with most of the electronic structure methods in \pyscf.
Furthermore, besides equilibrium solvation where the solvent polarization is governed
by the static electric susceptibility,
non-equilibrium solvation can also be treated within the framework of
TDDFT, in order to describe fast solvent response with respect to abrupt changes of 
the solute charge density. 
As an example, in Ref.~\onlinecite{Li2019}, the COSMO method was used to mimic
the protein environment of nitrogenase in electronic structure calculations for the P-cluster 
(Figure \ref{fig:solvent}).
For excited states generated by TDA, the polarizable embedding model\cite{Scheurer2019}
can also be used through an interface to the external library \textsc{CPPE}.\cite{Scheurer2019,cppe}

Currently, \pyscf provides some limited functionality for performing QM/MM  calculations
by adding classical point charges to the QM region. The implementation supports
all molecular electronic structure methods by decorating the underlying SCF methods.
In addition, MM charges can be used together with the X2C method and implicit solvent treatments. 

\subsection{Relativistic treatments}
\pyscf provides several ways to include relativistic effects. In the
framework of scalar Hamiltonians, spin-free X2C theory,\cite{Liu2009} scalar effective core
potentials\cite{Flores-Moreno2006} (ECP) and relativistic pseudo-potentials can all be used for all methods in calculations
of the energy, nuclear gradients and nuclear Hessians.
At the next level of relativistic approximations, \pyscf provides spin-orbit ECP integrals,
and one-body and two-body spin-orbit interactions from the Breit-Pauli Hamiltonian and
X2C Hamiltonian for the spin-orbit coupling effects.\cite{Mussard2018}
Two component Hamiltonians with the X2C one-electron approximation, and four-component
Dirac-Coulomb, Dirac-Coulomb-Gaunt, and Dirac-Coulomb-Breit Hamiltonians
are all supported in mean-field molecular calculations.

\subsection{MPI implementations}\label{sec:mpi}

\begin{figure}
\includegraphics{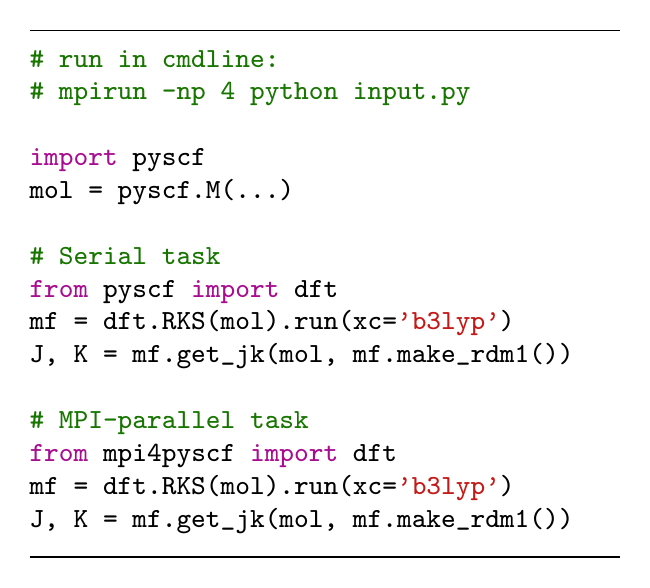}
\caption{Code snippet showing the similarity between serial and MPI-parallel DFT calculations.}
\label{code:mpi}
\end{figure}

\begin{figure}
  \begin{center}
    \includegraphics[width=.5\textwidth]{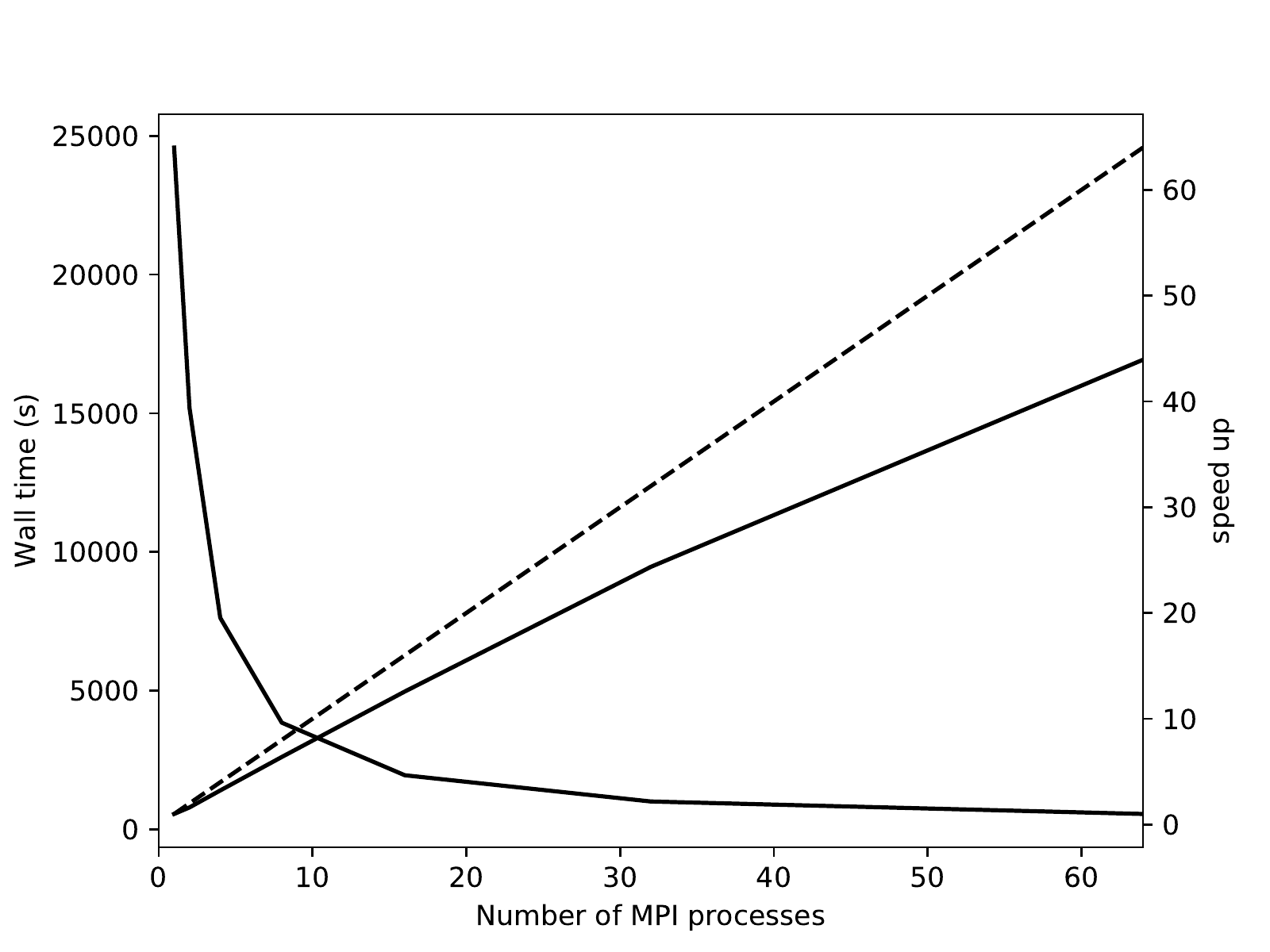}
  \end{center}
  \caption{Computation wall time of building the Fock matrix for the [H$_2$O]$_{512}$ cluster 
  	at the HF/VDZ level (12288 AO functions) using \pyscf's MPI implementation.
  	Each MPI process contains 32 OpenMP threads and 
    the speedup is compared to the single-node calculation with 32 OpenMP threads.}
  \label{fig:mpi}
\end{figure}

In \pyscf, distributed parallelism with MPI is implemented via an extension to the \pyscf main library known as \textsc{MPI4PySCF}.
The current MPI extension supports
the most common methods in quantum chemistry and crystalline material computations.
Table \ref{tb:mpi} lists the available MPI-parallel alternatives to the default serial (OpenMP)
implementations. The MPI-enabled modules implement almost identical APIs to the serial ones, allowing the same script to
be used for serial jobs and MPI-parallel jobs (Figure \ref{code:mpi}).
The efficiency of the MPI implementation is demonstrated in Figure \ref{fig:mpi}, which shows the wall time
and speedup of Fock builds for a system with 12,288 AOs with up to 64 MPI processes, each with 32 OpenMP threads.

To retain the simplicity of the \pyscf package structure, we use a
server-client mechanism to execute the MPI parallel code.
In particular, we use MPI to start the Python interpreter as a daemon that receives both the
functions and data on remote nodes.
When a parallel session is activated, the master process sends
the functions and data to the daemons.  The function object is decoded
remotely and then executed.
\red{For example, when building the Fock matrix in the PySCF MPI implementation,
the Fock-build function running on the master process first sends itself to
the Python interpreters running on the clients.
After the function is decoded on the clients, input variables (like the
density matrix) are distributed by the master process through MPI. Each client
evaluates a subset of the four-center two-electron integrals (with load balancing
performed among the clients) and constructs a partial Fock matrix, similarly to the Fock-build
functions in other MPI implementations. After sending the partial Fock matrices back to
the master process, the client suspends itself until it receives the next function.
The master process assembles the Fock matrices and then moves on to the next part of
the code.}
The above strategy is quite different from 
traditional MPI programs that hard-code MPI functionality into the code 
and initiate the MPI parallel context at the beginning of the program.
This \pyscf design brings the important benefit of being able to switch on and off MPI parallelism freely in the program
without the need to be aware of the MPI-parallel context. See Ref.~\onlinecite{Sun2018} for a more
detailed discussion of \pyscf MPI mode innovations.

\begin{table}
	\centering
	\begin{threeparttable}
	\caption{
		Methods with MPI support.  For solids, MPI support is currently provided only at the level
                of parallelization over $k$-points.  
	}
	\label{tb:mpi}
	\begin{tabular*}{.45\textwidth}{@{\extracolsep{\fill}}lcc}
                \hline\hline
Methods            & Molecules & Solids  \\
\hline
HF                 & Yes      & Yes      \\
DFT                & Yes      & Yes      \\
MP2                & Yes\tnote{a}  & Yes      \\
CCSD               & Yes\tnote{a}  & Yes      \\
		\hline\hline
	\end{tabular*}
	\begin{tablenotes}
		\item[a] closed shell systems only
	\end{tablenotes}
	\end{threeparttable}
\end{table}

\section{The \pyscf simulation ecosystem} \label{sec:extensions}

\pyscf is widely used as a development tool, and many groups have developed and made available their own projects that
either interface to \pyscf or can be used in a tightly coupled manner to access greater functionality.
We provide a few examples of the growing \pyscf ecosystem below, which we separate into use cases:
(1) external projects to which \pyscf provides and maintains a native interface, and (2) external projects 
that build on \pyscf.

\subsection{External projects with native interfaces}

\pyscf currently maintains a few native interfaces to external projects, including:
\begin{itemize}
  \item \textsc{geomeTRIC}\cite{Wang2016} and \textsc{pyberny}.\cite{pyberny} 
    These two libraries
    provide the capability to perform geometry optimization and interfaces to them are
    provided in the \pyscf GEOMOPT module. As shown in Figure \ref{code:geomopt},
    given a method that provides energies and nuclear gradients, the geometry
    optimization module generates an object that can then be used by these external optimization libraries.

  \item \textsc{DFTD3}.\cite{Grimme2010,libdftd3} This interface allows to add the DFTD3\cite{Grimme2010} correction to the
    total ground state energy as well as to 
    the nuclear gradients in geometry optimizations.

  \item DMRG, SHCI, and FCIQMC programs (\textsc{Block},\cite{Sharma2012} 
    \textsc{CheMPS2},\cite{Wouters2014}
    \textsc{Dice},\cite{Holmes2016,Sharma2017,Smith2017}
    \textsc{Arrow},\cite{Holmes2016,Sharma2017,Li2018}
    and \textsc{NECI}\cite{Booth2014}). 
    These interfaces closely follow the conventions of \pyscf's FCI 
    module. As such, they can be used to replace the FCI 
    solver in MCSCF methods (CASCI and CASSCF) to study large active 
    space multi-reference problems.
  
  \item \textsc{Libxc}\cite{Lehtola2018} and \textsc{XCFun}.\cite{Ekstrom2010}  
    These two libraries are tightly integrated into the \pyscf code.
    While the \pyscf DFT module allows the user to customize exchange correlation (XC) functionals by 
    linearly combining different functionals, the individual XC functionals and 
    their derivatives are evaluated within these libraries.
  
  \item \textsc{TBLIS}.\cite{Matthews2016,Huang2017a,tblis} The tensor contraction library TBLIS offers similar functionality
    to the \textsf{numpy.einsum} function while delivering substantial speedups.
    Unlike the BLAS-based ``transpose-GEMM-transpose'' scheme which
    involves a high memory footprint due to the transposed tensor
    intermediates, TBLIS achieves optimal tensor contraction performance
    without such memory overhead. 
    The TBLIS interface in \pyscf provides an \textsf{einsum}
    function which implements the \textsf{numpy.einsum} API but
    with the TBLIS library as the
    contraction back-end.

  \item \textsc{CPPE}.\cite{Scheurer2019,cppe} This library provides a polarizable embedding solvent model and can be integrated into \pyscf calculations for
   \red{ground-state mean-field and 
  post-SCF methods. In addition, an interface to TDA is currently supported for excited-state calculations.}
\end{itemize}

\subsection{External projects that build on \pyscf}

There are many examples in the literature of quantum chemistry and electronic structure simulation
packages that build on \pyscf. The list below is by no means exhaustive, but gives an idea of the range of projects
using \pyscf today.
\begin{enumerate}
  \item \emph{Quantum Monte Carlo}.
Several quantum Monte Carlo programs, such as
\textsc{QMCPACK},\cite{Kim2018}
\textsc{pyQMC},\cite{pyqmc}
\textsc{QWalk},\cite{Wagner2009} and
\textsc{HANDE}\cite{Spencer2019a}
support reading  wavefunctions and/or Hamiltonians generated by \pyscf.
In the case of \textsc{pyQMC}, \pyscf is integrated as a dependent module.

\item \emph{Quantum embedding packages}. Many flavours of quantum embedding, including density matrix
  embedding and dynamical mean-field theory, have been implemented on top of \pyscf.
  Examples of such packages include \textsc{QSoME},\cite{Chulhai2017,Petras2019,Goodpaster2019} 
  \textsc{pDMET},\cite{Hermes2019, Pham2020} 
  \textsc{PyDMFET},\cite{Zhang2018a}
  \textsc{Potato},\cite{Cui2019,Zhu2020}
  and the \textsc{OpenQEMIST} package,\cite{Yamazaki2018} 
  which all use \pyscf to manipulate wavefunctions and embedding Hamiltonians and to provide many-electron solvers. 

\item \emph{General quantum chemistry}. \pyscf can be found as a component of tools developed 
  for many different kinds of calculations, including localized active space self-consistent field (LASSCF),\cite{Hermes2019}
  multiconfiguration pair-density functional theory (MC-PDFT),\cite{Gagliardi2017} 
  and state-averaged CASSCF energy and analytical
gradient evaluation (these all use the \pyscf MCSCF module to optimize multi-reference
wavefunctions), as well as for localized orbital construction via the \textsc{Pywannier90} library.\cite{Pham2020}
The \textsc{PyMBE} package,\cite{pymbe} which implements the many-body expanded full CI 
method,\cite{Eriksen2017,Eriksen2018,Eriksen2019,Eriksen2019a}
utilizes \pyscf to perform all the underlying electronic structure calculations.
Green's functions methods such as the second-order Green's function theory (GF2) 
and the self-consistent GW approximation have been explored
using \pyscf as the underlying ab initio infrastructure.\cite{Iskakov2019}
In the linear scaling program LSQC,\cite{Li2015,Li2016}
\pyscf is used to generate reference wavefunctions and integrals for the cluster-in-molecule local correlation method.
The \textsc{APDFT} (alchemical perturbation density functional theory) program\cite{VonRudorff2018,VonRudorff2019}
interfaces to \pyscf for QM calculations.
In the \textsc{PySCF-NAO} project,\cite{Koval2019}
large-scale ground-state and excited-state methods are implemented based on 
additional support for numerical atomic orbitals, which has been integrated into an active branch of \pyscf.
The \textsc{PyFLOSIC} package\cite{Schwalbe2019} evaluates self-interaction
corrections with Fermi-L\"owdin orbitals in conjunction with the \pyscf DFT module.
Further, \pyscf FCI capabilities are used in the \textsc{molsturm}
package\cite{Herbst2018} for the development of Coulomb Sturmian basis functions,
and \pyscf post-HF methods appear in \textsc{VeloxChem}\cite{Rinkevicius2019} 
and \textsc{adcc}\cite{Herbst2020} for spectroscopic and excited-state simulations.

\end{enumerate}

\section{Beyond electronic structure}

\subsection{\pyscf in the materials genome initiative and machine learning} 

As discussed in section~\ref{sec:intro}, one of our objectives when developing \pyscf was to create a tool which could be used
by non-specialist researchers in other fields. With the integration 
of machine learning techniques into molecular and materials simulations, we find that
\pyscf is being used in many applications in conjunction with machine learning.
For example, the flexibility of the \pyscf DFT module has allowed it to be used to test exchange-correlation
functionals generated by machine-learning
protocols in several projects,\cite{Dick2019,Ji2018}
and has been integrated into other machine learning workflows.\cite{Hermann2019,Han2019}
\pyscf can be used as a large-scale computational engine for quantum chemistry data generation.\cite{Chen2019,Lu2019}
Also, in the context of machine learning of wavefunctions, \pyscf has been used as the starting point to
develop neural network based approaches for SCF initial guesses,\cite{scfinitguess}
for the learning of HF orbitals by the DeepMind team,\cite{Pfau2019} and
for Hamiltonian integrals used by fermionic neural nets in \textsc{NetKet}.\cite{Choo2019}

\subsection{\pyscf in quantum information science}

Another area where \pyscf has been rapidly adopted as a development tool
is in the area of quantum information science and quantum computing.
This is likely because Python is the de-facto standard programming language in the 
quantum computing community.
For example, \pyscf is one of the
standard prerequisites to carry out molecular simulations in the \textsc{OpenFermion}\cite{McClean2017} library, the
\textsc{QisKit-Aqua}\cite{qiskit} library and the
\textsc{OpenQEMIST}\cite{Yamazaki2018} package.
Via \pyscf's GitHub page, we see a rapidly increasing number
of quantum information projects which include \pyscf as a program dependency. 


\section{Outlook}

After five years of development, the \pyscf project can probably now be considered to be a feature complete and mature
tool. Although no single package can be optimal for all tasks, we believe \pyscf to a large extent meets its original
development criteria of forming a library that is not
only useful in simulations but also in enabling the customization and development of new electronic structure methods.

With the recent release of version 1.7, the current year marks the end of development of the version 1 branch of \pyscf.
As we look towards \pyscf version 2, we expect to build additional innovations, for example, in the areas of faster electronic structure methods
for very large systems, further support and integration for machine learning and quantum computing applications,
better integration of high-performance computing libraries and more parallel implementations, and perhaps even forays
into dynamics and classical simulations.
\red{Beyond feature development, we will expand our efforts in documentation and in quality assurance and testing}.
We expect the directions of implementation to continue to be guided by and organically
grow out of the established \pyscf ecosystem. However, regardless of the scientific directions and methods implemented within \pyscf,
the guiding philosophy described in this article will continue to lie at the heart of \pyscf's development. We believe these guiding principles will
help ensure that \pyscf remains a powerful and useful tool in the community for many years to come.

\section*{Data Availability Statement}
The data that supports the findings of this study are available within the article,
and/or from the corresponding author upon reasonable request.

\begin{acknowledgements}
  As a large package, the development of \pyscf has been supported by different
  sources. Support from the US National Science Foundation via award no. 1931258
  (T.C.B., G.K-L.C., and L.K.W.) is acknowledged to integrate high-performance
  parallel infrastructure and faster mean-field methods into \pyscf.
  Support from the US National Science Foundation via award no. 1657286
(G.K.-L.C.) and award no. 1848369 (T.C.B.) is acknowledged for various aspects of the development
of many-electron wavefunction methods with periodic boundary conditions. Support
for integrating \pyscf into quantum computing platforms
  is provided partially by the Department of Energy via award no. 19374 (G.K.-L.C). The Simons Foundation is gratefully acknowledged
  for providing additional support for the continued maintenance and development of \pyscf.   The Flatiron Institute is a division of the Simons Foundation.
  M.B. acknowledges support from the Departemento de Educación of the Basque Government through a PhD grant, as well as from Euskampus and the DIPC at the initial stages of his work.
  J.C. is supported by the Center for Molecular Magnetic Quantum Materials (M2QM), an Energy Frontier Research Center funded by the US Department of Energy, Office of Science, Basic Energy Sciences under Award DE-SC0019330.
  J.J.E. acknowledges financial support from the Alexander von Humboldt Foundation and the Independent Research Fund Denmark.
  M.R.H. and H.Q.P. were partially supported by the U.S. Department of Energy, Office of Science, Basic Energy Sciences, Division of Chemical Sciences, Geosciences and Biosciences under Award \#DE-FG02-17ER16362, while working in the group of Laura Gagliardi at the University of Minnesota.
  P.K. acknowledges financial support from the Fellows Gipuzkoa program of the Gipuzkoako Foru Aldundia through the FEDER funding scheme of the European Union.
  S.L. has been supported by the Academy of Finland (Suomen Akatemia) through project number 311149.
  A.P. thanks Swiss NSF for the support provided through the Early Postdoc. Mobility program (project P2ELP2\_175281).
  H.F.S acknowledges the financial support from the European Union via Marie Skłodowska-Curie Grant Agreement No. 754388 and LMUexcellent within the German Excellence Initiative (No. ZUK22).
  S.B. and J.E.T.S. gratefully acknowledge support from a fellowship through The Molecular Sciences Software Institute under NSF Grant ACI-1547580. 
  S.S. acknowledges support of NSF grant CHE-1800584.
  S.U. acknowledges the support of NSF grant CHE-1762337. 
  The National Science Foundation Graduate Research Fellowship Program is acknowledged for support of J.M.Y.
\end{acknowledgements}



\bibliographystyle{aipnum4-1}
\bibliography{ref}

\end{document}